\documentclass[useAMS,usenatbib,referee]{mn2e}
\usepackage[dvips]{graphicx}

\def\lesssim{\mathrel{\hbox{\rlap{\hbox{\lower4pt\hbox{$\sim$}}}\hbox{$<$}}}}
\def\gtrsim{\mathrel{\hbox{\rlap{\hbox{\lower4pt\hbox{$\sim$}}}\hbox{$>$}}}}
\newcommand{\mincir}{\raise
-2.truept\hbox{\rlap{\hbox{$\sim$}}\raise5.truept
\hbox{$<$}\ }}
\newcommand{\magcir}{\raise
-2.truept\hbox{\rlap{\hbox{$\sim$}}\raise5.truept
\hbox{$>$}\ }}
\newcommand{\siml}{\raise -2.truept\hbox{\rlap{\hbox{$\sim$}}\raise5.truept
\hbox{$<$}\ }}
\newcommand{\simg}{\raise -2.truept\hbox{\rlap{\hbox{$\sim$}}\raise5.truept
\hbox{$>$}\ }}
\newcommand{\be}{\begin{equation}}
\newcommand{\ee}{\end{equation}}
\newcommand{\ba}{\begin{eqnarray}}
\newcommand{\ea}{\end{eqnarray}}

\newcommand {\kss} {km~s$^{-1}$}
\newcommand {\msun} {$\mathrm{h^{-1}_{70} \  M_{\odot} \;}$}

\title[Global Properties of ABCG\,209] 
{Global Properties of the Rich Cluster ABCG\,209 at
z$\sim$0.2. Spectroscopic and Photometric Catalogue\footnote{Based on
observations collected at European Southern Observatory (ESO ObsID.s
068.A-0116, 074.A-0073 and 075.A-0845) and at the Telescopio Nazionale
Galileo (TNG ObsID AOT12/TAC\_07).}\thanks{Tables 2 and 3 are
available only in electronic format.}}

\author[A. Mercurio et al.]  {A.~Mercurio$^1$,
F.~La Barbera$^1$, C. P.~Haines$^{1,2}$, P.~Merluzzi$^1$,
G.~Busarello$^1$ \newauthor 
and M.~Capaccioli$^{3,4}$\\
   $^1$INAF-Osservatorio Astronomico di Capodimonte, I-80131 Napoli\\
   $^2$School of Physics and Astronomy, University of Birmingham, Edgbaston,
Birmingham, B15 2TT, UK\\
$^3$Dipartimento di fisica, Universit\`a `Federico II' di
   Napoli, I-80126 Napoli\\
   $^4$INAF-VSTCen, I-80131 Napoli}

\date{Released 2006 Xxxxx XX}

\def\LaTeX{L\kern-.36em\raise.3ex\hbox{a}\kern-.15em
    T\kern-.1667em\lower.7ex\hbox{E}\kern-.125emX}

\begin{document}

\label{firstpage}

\maketitle

\begin{abstract}

This paper is aimed at giving an overview of the global properties of
the rich cluster of galaxies ABCG\,209. This is achieved by
complementing the already available data with new medium resolution
spectroscopy and NIR photometry which allow us to i) analyse in detail
the cluster dynamics, distinguishing among galaxies belonging to
different substructures and deriving their individual velocity
distributions, using a total sample of 148 galaxies in the cluster
region, of which 134 belonging to the cluster; ii) derive the cluster
NIR luminosity function; iii) study the Kormendy relation and the
photometric plane of cluster early-type galaxies (ETGs).  Finally we
provide an extensive photometric (optical and NIR) and spectroscopic
dataset for such a complex system to be used in further analyses
investigating the nature, formation and evolution of rich clusters of
galaxies.

The present study shows that the cluster is characterised by a very
high value of the line-of-sight velocity dispersion:
${\rm\sigma_v=1268^{+93}_{-84}}$ \kss, that results in a virial mass
of $M_{vir}=2.95^{+0.80}_{-0.78}\times 10^{15} $ \msun within
$R_{vir}=3.42 h^{-1}_{70}$ Mpc.  The analysis of the velocity
dispersion profile shows that such high value of $\sigma_v$ is already
reached in the central cluster region. There is evidence of three
significant substructures, the primary one having a velocity
dispersion of ${\rm\sigma_v=847^{+52}_{-49}}$ \kss, which makes it
consistent with mass estimates from weak lensing analyses.

This observational scenario confirms that ABCG\,209 is presently
undergoing strong dynamical evolution with the merging of two or more
subclumps. This interpretation is also supported by the detection of a
radio halo (Giovannini et al. \citeyear{gio06}) suggesting that there
is a recent or ongoing merging.

Cluster ETGs follow a Kormendy relation whose slope is consistent with
previous studies both at optical and NIR wavelengths.  We investigate
the origin of the intrinsic scatter of the photometric plane due to
trends of stellar populations, using line indices as indicators of
age, metallicity and $\alpha$/Fe enhancement. We find that the
chemical evolution of galaxies could be responsible for the intrinsic
dispersion of the Photometric Plane.

\end{abstract}


\section{Introduction}

Galaxy clusters are complex systems involving a variety of interacting
components: galaxies, hot and cold gas, dark matter. Among them, about
50\% are unrelaxed systems (Smith et al. \citeyear{smi05}) and a large
fraction contain substructures (e.g., Girardi et al.
\citeyear{gir06} and references therein), suggesting that they are
evolving via merging processes from poor groups to rich structures.
In fact, in hierarchical clustering cosmological scenarios, galaxy
clusters form from the accretion of subunits. Numerical simulations
show that clusters form preferentially through the anisotropic accretion
of subclusters along filaments (e.g., Colberg et al. \citeyear{col99},
Diaferio et al. \citeyear{dia01}).

Detailed multi-band studies of such systems are crucial to probe
structure formation scenarios as well as to investigate galaxy
evolution processes. ABCG\,209 is an ideal target for such a study,
since it is a rich (richness class \mbox{${\rm R}=3$}; Abell et al.
\citeyear{abe89}), X-ray luminous (\mbox{${\rm L}_{{\rm
X}}$(0.1--2.4\,keV$)\sim2.7\times10^{45}\,h^{-2}_{70}\,$erg\,s$^{-1}$},
Ebeling et al. \citeyear{ebe96}; \mbox{${\rm T}_{{\rm
X}}\sim10$\,keV}, Rizza et al. \citeyear{riz98}), and massive cluster
(\mbox{${\rm M(R}<{\rm R}_{{\rm
vir}})=2.$3--3.1$\times10^{15}h^{-1}_{70}{\rm M}_{\odot}$} Mercurio et
al.  \citeyear{mer03a}, Paulin-Henriksson et
al. \citeyear{pau07}). Evidence for a complex dynamical status comes
from the X-ray emission: the hot gas is elongated and distributed
asymmetrically showing two main clumps (Rizza et
al. \citeyear{riz98}). However no strong cooling flow is
detected. Moreover, the young dynamical state is indicated by the
possible presence of a radio halo (Giovannini et al.
\citeyear{gio99}, \citeyear{gio06}), which has been suggested to be
the result of a recent cluster merger, through the acceleration of
relativistic particles by the merger shocks (Feretti \citeyear{fer02},
\citeyear{fer07}).

Dynamical properties of the cluster were analysed through a
spectroscopic survey of 112 cluster members (Mercurio et
al. \citeyear{mer03a}) and photometric properties of the cluster
galaxies were derived out to radii of 3-4 $h_{70}^{-1}$ Mpc, taking
advantage of wide field CFHT $B$- and $R$-band images (Haines et
al. \citeyear{hai04}). Our analyses (Mercurio et
al. \citeyear{mer03a}, \citeyear{mer03b}, \citeyear{mer04}; La Barbera
et al. \citeyear{lab03a}, \citeyear{lab03b}, \citeyear{lab04}; Haines
et al. \citeyear{hai04}) showed that ABCG\,209 has a very complex
structure, with: {\bf a)} a high value of the line-of-sight velocity
dispersion, with
\mbox{$\sigma_{v}=1394^{+88}_{-99}$\,km\,s$^{-1}$}; {\bf b)} a
significantly non-Gaussian redshift distribution, and a velocity
gradient (and elongation of the cD galaxy) in the SE-NW direction;
{\bf c)} the presence of substructures in the X-ray emission; {\bf d)}
the presence of a merger clump observed 1-2 Gyr after the merging;
{\bf e)} a strong spatial and spectral segregation of galaxies, with
(i) young blue emission line galaxies uniformly distributed in the low
density regions, (ii) blue post-starburst galaxies aligned in a
direction perpendicular to the cluster elongation, and (iii) red
post-starburst galaxies and early type galaxies distributed along the
cluster elongation.  Moreover, we have examined the effect of cluster
environment, as measured in terms of the local surface density of
$R<$23.0 mag galaxies, on the global properties of the cluster
galaxies, through their luminosity functions, colour-magnitude
relations, and average colours. The LFs for galaxies within the
virialized region are found to be well described by single Schechter
functions, although there is an indication of a dip at $R$=20-20.5
mag.  The faint-end slope shows a strong dependence on environment,
becoming steeper at more than 3$\sigma$ significance level from high-
to low-density environments.  We explain this trend as a combination
of the morphology-density relation and dwarf galaxies being
cannibalised and disrupted by the cD galaxy and interactions with the
intra-cluster medium in the cluster core.

Our weak lensing analysis (Paulin-Henriksson et al. \citeyear{pau07})
confirms that ABCG\,209 is a massive cluster, although the mass
estimated by weak lensing (${\rm M(R}<{\rm R}_{{\rm
200}})=7.7_{-2.7}^{+4.3} \times10^{14}h^{-1}_{70}{\rm M}_{\odot}$)
is lower than that obtained by Mercurio et al. (\citeyear{mer03a})
from the analysis of the dynamical properties. The centres of the
X-ray emission, dark matter and galaxy distributions all appear offset
from one another, with the centre of mass found from the weak lensing
analysis lying between that of the X-ray and galaxy distributions,
with all three centres of mass aligned on the main SE-NW axis of the
cluster (see Figs. 6 and 7 of Paulin-Henriksson et
al. \citeyear{pau07}). Such an effect is seen for the more extreme
``Bullet cluster'' (Clowe et al. \citeyear{clo04}), and seems to
reflect the different responses of the gas and dark matter components
to the merger, in agreement with the merging scenario for ABCG\,209.

The above results show that it is crucial to relate the properties of
member galaxies to the global properties of clusters, such as mass
and dynamical state. The spatial and kinematical analysis of member
galaxies allows to detect and measure the amount of substructure
and to identify and analyse possible pre-merging clumps or merger
remnants. In addition, subclustering is important for processes of
galaxy evolution. In fact, although the high galaxy velocity dispersion
associated with a relaxed, virialized clusters inhibits galaxy
mergers, in subclusters, where the velocity dispersions are lower,
such mergers are more probable and could have an important role in
building elliptical galaxies (see Moss \citeyear{mos06}).

Identifying substructures is difficult with galaxy positions alone,
since projected distribution of galaxy clusters could show clumps that
are not real. The optical spectroscopy of member galaxies is the most
reliable tool to investigate these clumps and allows us to study the
dynamics of cluster mergers, since it provides direct information on
the velocity field. However, this is often an arduous investigation
due to the limited number of galaxies usually available to trace the
velocity field. 

In order to achieve the resolution needed to understand the complex
dynamics of ABCG\,209 and to investigate the environmental effects on
galaxy evolution, we complement the previously described dataset with
medium resolution spectra analysing a total sample of 148 galaxies in
the cluster region. The enlarged data sets allow us to better
distinguish among galaxies belonging to different substructures and to
derive their individual velocity distributions. A detailed study of
substructures is also important to investigate the reason for the
discrepancy between lensing and dynamical mass estimates. We also
present new $K$-band observations, that allow us to derive the cluster
NIR luminosity function (LF) and are directly related to the galaxy
mass function. Additionally, at infrared wavelengths the galaxy
luminosities do not depend strongly on the details of their stellar
populations (Gavazzi, Pierini, \& Boselli \citeyear{gav96}).

In this paper we provide the spectroscopic catalogue of the new
targets and a photometric catalogue with measured photometric
redshifts, $B$-, $V$-, $R$- and $K$-band total magnitudes and
spectroscopic redshifts when available.  We also derive structural
parameters of cluster galaxies, selected according to the photometric
and spectroscopic redshifts; and we investigate the photometric
correlations of ETGs, such as the mean surface
brightness-size relation, also known as the kormendy relation (KR;
Kormendy \citeyear{kor77}) and the Photometric Plane (PHP; La Barbera
et al.  \citeyear{lab05}).  

The paper is organised as follows. We present the new data in Sect. 2
and we derive photometric and spectroscopic redshifts in Sect. 3 where
the catalogue of the cluster members is also described.  We analyse
the global cluster properties in Sect. 4 and the internal dynamics and
substructures in Sect. 5. We describe the derivation of structural
properties of ETGs in Sect. 6, deriving also the KR and the PHP.
Finally we discuss and summarise our results in Sect. 7.

Unless otherwise stated, we give errors at the 68\% confidence level
(hereafter c.l.). Throughout the paper, we assume a flat cosmology
with $\Omega_{\rm m}=0.3$, $\Omega_{\Lambda}=0.7$, and $H_0$=70 km
s$^{-1} \;$ Mpc$^{-1}$. For this cosmological model, $1{\arcmin}$ corresponds
to 271 kpc at the cluster redshift.

\section{Observations and data reduction}
\label{sec:2}

Spectroscopic observations were carried out at the ESO New Technology
Telescope (NTT) with the ESO Multi Mode Instrument (EMMI) and at the
Telescopio Nazionale Galileo (TNG) with the Device Optimized for the
LOw RESolution (DOLORES), while NIR photometric data were collected
with the Son OF ISAAC (SOFI) at NTT.

\subsection{Photometry}
\label{sec:21}

New $K$-band images for the cluster of galaxies ABCG\,209 \, were
collected with SOFI operating in the LARGE FIELD observing mode,
providing a field of view of $5{\arcmin} \times 5{\arcmin}$ with a
pixel scale of $\rm 0.288{\arcsec}/pxl$.  Eight overlapping fields
were observed, each of them with a dithered sequence of 36 exposures.
We adopted a $\mathrm{20{\arcsec}}$ dithering box, with DIT and NDIT
values of $\mathrm{ 6~s}$ and $\mathrm{ 10 }$, respectively. This
resulted in an integration time of $60$s for each exposure, and a total
integration time of $2160$s for each field.  Sky conditions were
photometric during five out of six observing nights.  During those
nights we observed standard stars from the list of~\citet{PMK98}, at
five different positions on the chip.

\begin{figure*}
\center
\includegraphics[width=0.8\textwidth]{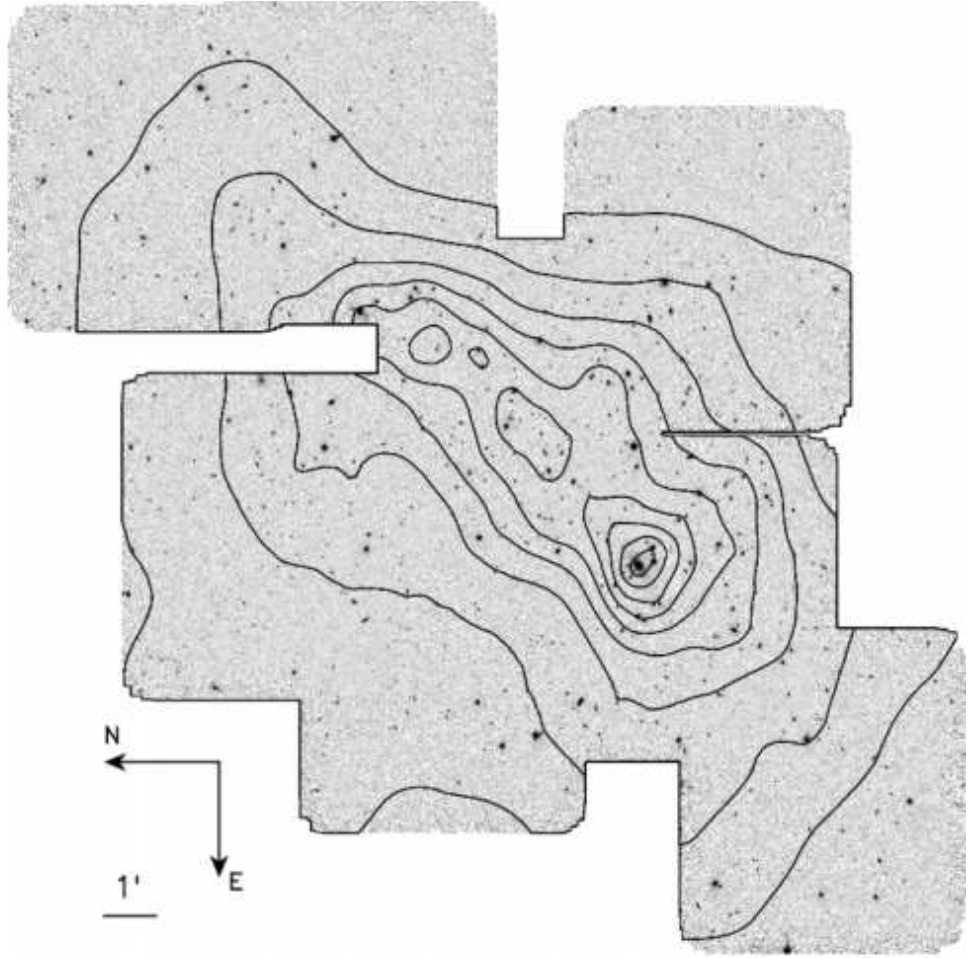}
\caption{K-band mosaic image of ABCG\,209.  Contours are obtained from
  the projected number density map of $\mathrm{K}<17.5$ mag
  galaxies. The contour density values are equally spaced by 1.27
  galaxies arcmin$^{-2}$, with the outermost and innermost contours
  corresponding to surface density values of 1.4 galaxies
  arcmin$^{-2}$ and 9 galaxies arcmin$^{-2}$, respectively.  The
  mosaic covers an area of $\sim 0.0525$ square degrees, corresponding
  to $\sim 1.39$ h$_{70}^{-2}$ Mpc$^2$ at z=0.209.}
\label{MOSAICO}
\end{figure*}

The data were reduced using {\sc fortran} routines developed by the
authors.  For each dithering sequence, the exposures were dark
subtracted and flat-field corrected, using a superflat frame obtained
by median combining all the images taken during a given night.  After
this procedure, the magnitudes of the standard stars showed a rms
variation across the chip of $\rm \sim 0.05$ mag.  To achieve a better
accuracy, we retrieved $K$-band images from the ESO archive for a
standard star observed at different positions in a $5 \times 5$ grid
on the SOFI frame.  We chose standard star observations as close as
possible in time to those of ABCG\,209.  Illumination correction
frames were then obtained for each night by measuring the magnitude of
the standard star as a function of the position on the frame. This
allowed the low spatial frequency component of the flat-field to be
corrected to better than $1 \%$.  Since sky subtraction is a very
troublesome step for the reduction of NIR data, particularly in high
density regions, this point was carefully dealt with by a two step
procedure.  First, each exposure was sky subtracted by computing the
sky frame from the median of the six closest frames along the
sequence.  The images were then registered with integer shifts and
combined using a sigma clipping algorithm for cosmic ray rejection.
This procedure alone overestimates the sky level in the extended halos
of galaxies.  To minimise this effect, we reiterated the sky
subtraction as follows.  The initial combined images were used to
obtain mask frames for the sources in the field, by running SExtractor
with the checkimage OBJECTS option.  For each sequence, the mask was
expanded in order to 'cover' the galaxy halos and was de-registered to
each dithered exposure. The hot and bad pixels were also masked.  The
sky frames were estimated as the average of the six closest frames to
the exposures from that sequence, rejecting masked pixels.  The
exposures of each dithering sequence were then sky subtracted and
combined with the {\sc iraf}\footnote{IRAF is distributed by the
National Optical Astronomy Observatories, which are operated by the
Association of Universities for Research in Astronomy, Inc., under
cooperative agreement with the National Science Foundation.} task {\sc
imcombine}. The images of the different fields were combined by taking
into account their different zero-points, resulting in a final
$K$-band mosaic with an average seeing FWHM of $\sim0.8{\arcsec}$. The
mosaic is shown in Fig.~\ref{MOSAICO}.

The photometric calibration was performed for the $K_s$ standard filter
\citep{PMK98}, deriving the instrumental magnitudes of the standard
stars within an aperture of $8{\arcsec}$ diameter. The airmass
correction was performed by using an extinction coefficient
$\mathrm{A_K }$ = 0.05~mag/airmass, which was derived by comparing the
magnitudes of bright objects in the field for different airmasses.
The typical accuracy on the zero-point of each photometric night
amounts to $\sim 0.01$ mag. The images of the different fields were
normalised to the average zero-point value of $22.425$ (scaled to $\rm
1~s$ exposure time).

\subsection{Spectroscopy}
\label{sec:22}

The spectroscopic data were obtained with the multi-object spectroscopy
(MOS) mode. Targets were selected according to the $B$-$R$ vs. $R$
colour-magnitude relation (CM). We gave priority to the galaxies lying
on the CM relation up to $R$=20.0 mag. We acquired one mask with EMMI (field
of view $5^\prime \times 8.6^\prime$), centred on
$\alpha_{J2000}=$01:31:53.0, $\delta_{J2000}=$-13:37:42.8, allocating
28 slits and one with DOLORES (field of view $6^\prime \times
7.7^\prime$), centred on $\alpha_{J2000}=$01:31:52.0,
$\delta_{J2000}=$-13:37:30.8, allocating 33 slits. We integrated for
8100 sec with EMMI-Grism\#5 yielding a dispersion of $\sim$1.6 \AA/pxl
(resolution=$\sim$4.8 \AA \ \ FWHM) in the spectral range 380-702 nm, and
for 16200 sec with DOLORES-grism MR-B, giving a dispersion $\sim$1.9
\AA/pxl (resolution=$\sim$4.2 \AA \ \ FWHM) in the spectral range
350-700 nm.

Each scientific exposure (as well as flat fields and calibration
lamps) was bias subtracted. The individual spectra were extracted and
flat field corrected. Cosmic rays were rejected in two steps. First,
we removed the cosmic rays lying close to the objects by interpolation
between adjacent pixels, then we combined the different exposures by
using the {\sc iraf} task {\sc imcombine} with the algorithm {\sc
crreject} (the positions of the objects in different exposures were
checked before). Wavelength calibration was obtained using He-Ar and
He lamp spectra for EMMI and DOLORES, respectively. The typical rms
scatter around the dispersion relation was $\sim$ 8 km s$^{-1}$
and 11 km s$^{-1}$
for EMMI and DOLORES respectively.

The positions of the objects in the slits were defined interactively
using the {\sc iraf} package {\sc apextract}. The exact object
position within the slit was traced in the dispersion direction and
fitted with a low order polynomial to allow for atmospheric
refraction. The spectra were then sky subtracted and the rows
containing the object were averaged to produce the one-dimensional
spectra.  The signal-to-noise ratio per pixel of the one-dimensional
galaxy spectra belonging to the cluster ranges from about 6 to 18 in
the region around the Mg line (at 625 nm in the spectra at z$\sim$0.2)
for EMMI spectra and from about 8 to 27 for the spectra obtained with
DOLORES.

\section{Redshift measurements}
\label{sec:3}

\subsection{Spectroscopic redshifts}
\label{sec:31}

Redshifts were derived using the cross-correlation technique
\citep{ton81}, as implemented in the {\sc rvsao} package. We adopted
galaxy spectral templates from \citet{ken92}, corresponding to
morphological types E, S0, Sa, Sb, Sc and Ir. The correlation was
computed in the Fourier domain. We define the redshift as the value
given by the {\it best template spectra}, i.e. the template
producing the highest value of the correlation $R-value$ given by
{\sc rvsao} that gives an indication of the signal-to-noise of the
correlation peak.

Of the 61 observed spectra, 44 
turned out to be at the redshift of ABCG\,209, 1  
is a foreground galaxy and 6 are background galaxies. 
In 10 
cases we could not determine the redshift. In Table
\ref{spec_catalogue} we report spectroscopic redshift measurements of
the observed galaxies.

\begin{table}
        \caption[]{Spectroscopic data. Running number for
         galaxies in the present sample, right ascension 
	 and declination (Col.~1, Col.~2); heliocentric
         corrected redshift $\mathrm{z}$ (Col.~3).}
         \label{spec_catalogue}
           $$ 
           \begin{array}{c c c}
            \hline
            \noalign{\smallskip}
            \hline
            \noalign{\smallskip}

\mathrm{\alpha} & \mathrm{\delta}  & \mathrm{z}\\
            \hline
            \noalign{\smallskip}   

 01\ 31\ 45.33 & -13\ 41\ 53.0  & 0.2004\pm0.0004  \\
 01\ 31\ 56.81 & -13\ 41\ 31.7  & 0.2182\pm0.0002  \\
 01\ 31\ 47.02 & -13\ 41\ 17.9  & 0.2030\pm0.0002  \\
 01\ 31\ 48.41 & -13\ 41\ 00.4  & 0.2079\pm0.0002  \\
 01\ 31\ 45.78 & -13\ 40\ 37.8  & 0.2031\pm0.0002  \\
 01\ 32\ 01.00 & -13\ 40\ 17.5  & 0.2113\pm0.0003  \\
 01\ 31\ 52.41 & -13\ 40\ 02.0  & 0.1989\pm0.0003  \\
 01\ 31\ 42.47 & -13\ 39\ 44.3  & 0.2539\pm0.0003^a\\
 01\ 32\ 01.27 & -13\ 39\ 33.3  & 0.2103\pm0.0005  \\
 01\ 31\ 57.49 & -13\ 39\ 23.4  & 0.2165\pm0.0003  \\ 
 01\ 31\ 47.06 & -13\ 39\ 06.7  & 0.3116\pm0.0003^a\\
 01\ 31\ 45.35 & -13\ 38\ 46.0  & 0.2522\pm0.0003^a\\
 01\ 31\ 46.73 & -13\ 38\ 29.2  & 0.2519\pm0.0003^a\\
 01\ 31\ 52.37 & -13\ 37\ 40.6  & 0.1980\pm0.0002  \\
 01\ 31\ 52.91 & -13\ 37\ 34.6  & 0.2150\pm0.0002  \\
 01\ 31\ 56.40 & -13\ 37\ 18.3  & 0.2080\pm0.0003  \\
 01\ 31\ 51.79 & -13\ 36\ 54.8  & 0.2074\pm0.0002  \\
 01\ 31\ 53.25 & -13\ 36\ 44.2  & 0.2098\pm0.0004  \\
 01\ 31\ 52.52 & -13\ 36\ 27.5  & 0.2110\pm0.0003  \\
 01\ 31\ 42.34 & -13\ 36\ 06.4  & 0.2080\pm0.0003  \\ 
 01\ 32\ 04.29 & -13\ 35\ 57.0  & 0.3702\pm0.0005^a\\
 01\ 31\ 49.42 & -13\ 35\ 44.5  & 0.2164\pm0.0004  \\
 01\ 31\ 44.94 & -13\ 35\ 11.3  & 0.2130\pm0.0002  \\
 01\ 31\ 49.39 & -13\ 34\ 59.3  & 0.2078\pm0.0003  \\  
 01\ 31\ 49.14 & -13\ 34\ 45.5  & 0.4025\pm0.0002^a\\
 01\ 31\ 48.95 & -13\ 34\ 34.9  & 0.2058\pm0.0005  \\
 01\ 31\ 48.33 & -13\ 34\ 23.5  & 0.1486\pm0.0004^b\\
 01\ 31\ 43.11 & -13\ 34\ 07.5  & 0.2109\pm0.0002  \\
 01\ 31\ 40.63 & -13\ 33\ 46.3  & 0.2097\pm0.0003  \\
              \noalign{\smallskip}			    
            \hline					    
            \noalign{\smallskip}			    
            \hline					    
         \end{array}
   $$ 
\begin{list}{}{}  
\item[$^{\mathrm{a}}$] Background galaxy.
\item[$^{\mathrm{b}}$] Foreground galaxy.
\end{list}
         \end{table}

In order to estimate the uncertainties in the redshift measurements,
we considered the error calculated with the cross-correlation
technique, which is based on the width of the peak and on the
amplitude of the antisymmetric noise from the cross correlation
(cf. manual of the {\sc xcsao} task). The wavelength calibration
errors (see Sect.\ref{sec:21}) turned out to be negligible in this
respect. The errors derived from the cross-correlation could however
be smaller than the true errors (e.g. Boschin et al. \citeyear{bos06}
and references therein).  We checked the error estimates by comparing
redshifts computed for 17 galaxies observed with EMMI and 5 with
DOLORES which were already acquired in a previous observation carried
out with EMMI \citep{mer03a}.  The two data sets agree with a
one-to-one relation, and a reasonable value of $\chi^2$ for the fit,
in particular for the spectra acquired with DOLORES, was obtained when
the errors derived from the cross-correlation were multiplied by a
correction factor $\sim$1.5. A similar correction was obtained by
Malumuth et al. (\citeyear{mal92}; 1.6), Bardelli et
al. (\citeyear{bar94}; 1.87), and Quintana et al. (\citeyear{qui00};
1.57).  Since we are comparing redshifts measured from spectra
acquired in different periods and with different instruments, the
multiplicative factor obtained through this comparison takes also into
account possible external errors.

\subsection{Photometric redshifts}
\label{sec:32}
The redshifts of galaxies without spectroscopic information were
estimated by means of the photometric redshift technique, using a
similar approach to that described in \citet{bus02} and
\citet{lab03c}.  Photometric redshifts were estimated according to the
Spectral Energy Distribution fitting method (see \citealt{mas01a,
mas01b}, and references therein), using the new $K$-band photometry,
as well as the already published $B$-, $V$-, and $R$-band photometry
of ABCG\,209.  The $B$- and $R$-band data, described in Haines et
al.~\citeyear{hai04}, are available on the entire field covered by the
new $K$-band photometry, while the $V$-band imaging (see Mercurio et
al.~\citeyear{mer03b}) covers a smaller region of $\sim160$ arcmin$^2$
centred around the cluster core.  Photometry of sources in the
$K$-band mosaic are obtained by means of SExtractor (Bertin \& Arnouts
\citeyear{ber96}), as detailed in Sec.~\ref{KBANDMAG}.  Galaxy colours
were measured within an aperture of $5{\arcsec}$ diameter. Since all
the $BVRK$ images have similar seeing FWHM values
($0.7{\arcsec}-0.9{\arcsec}$), no aperture corrections were applied to
galaxy colours.

Photometric redshifts were measured for the 642 galaxies, with
available photometry in $B$, $R$, and $K$ bands, brighter than $R=21$
mag: the magnitude limit for spectroscopic observation.  We looked for
redshifts in the range $z \in [0.0,1.0]$ with a step of 0.01, imposing
that at a given redshift galaxy templates would be younger than the
age of the universe in the adopted cosmology.  We used the GISSEL03
spectral code~\citep{BrC03} to produce galaxy templates with a Scalo
IMF and an exponential SFR, $\rm e^{-t/\tau}$.  The colours of E/S0,
Sa/Sb, and Sc/Sd spectra were modelled by choosing $\rm \tau = 1,4$
and $\rm 15~Gyr$, respectively, while early-type galaxies with
different metallicities were described by using E/S0 models with
$Z/Z_\odot$=0.2, 0.4, 1 and 2.5.  The differential dust extinction of
the Milky Way was included in the computation of model colours by
adopting the extinction curve of \citet{CAR} and a colour excess of
$\rm E(B-V)=0.019$~\citep{SFD98}.  The uncertainty on the photometric
redshift, $\rm \delta z$, was estimated by performing numerical
simulations, shifting galaxy colours according to their corresponding
uncertainties, and recomputing each time the photometric redshifts.

\begin{figure}
\center
\includegraphics[width=0.7\textwidth]{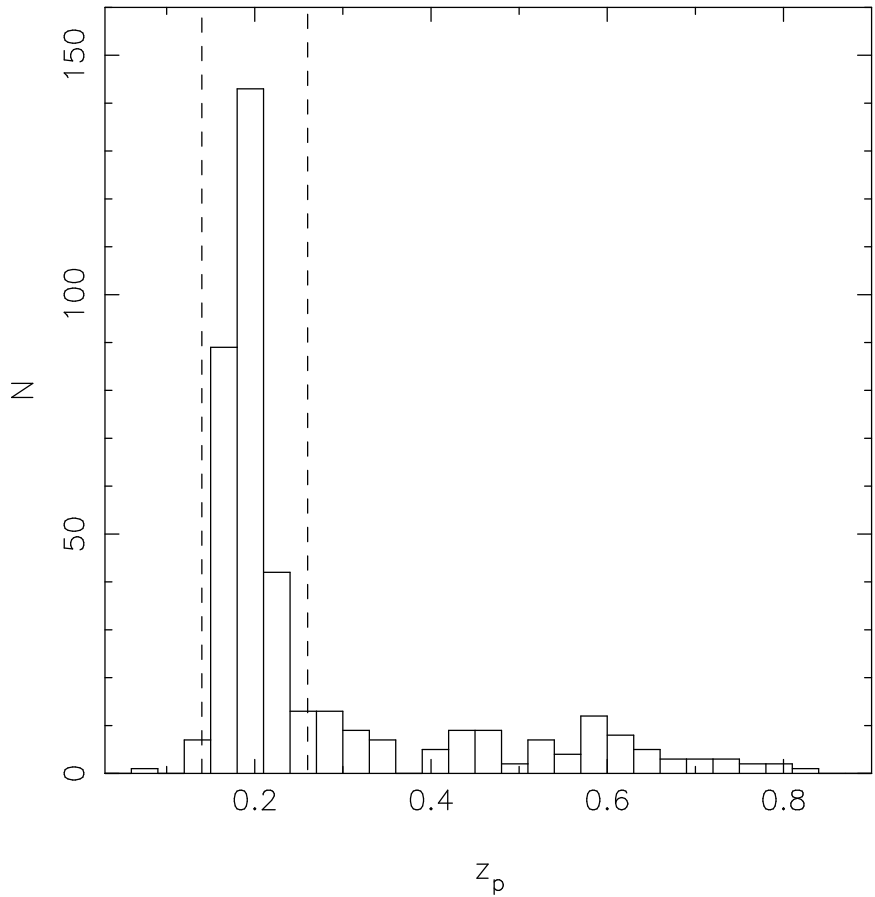}
\caption{Distribution of photometric redshifts for galaxies with a
relative uncertainty on photometric redshifts smaller than 0.07.
The dashed lines mark the redshift range that
defines likely cluster members.  }

\label{red_phot}
\end{figure}

We compared photometric and spectroscopic redshifts for $N=134$
galaxies in the cluster field. We found that galaxies with large
relative uncertainties on photometric redshifts were biased toward
higher redshift values.  In order to reduce this systematic effect, we
removed from the sample those galaxies with a value of $\delta z /
1+z$ larger than 0.07.  This selection leads to a sample of 399 (out
of 642) galaxies with reliable photometric redshift estimates (see
Table~\ref{phot_cat}).  As shown in Fig.~\ref{red_phot}, the
distribution of these photometric redshifts is dominated by the peak
around $z \sim 0.2$, indicating that most galaxies with $R \le 21$ in
the $K$-band field are actually cluster members.  After comparison
with spectroscopic redshifts, we chose to mark as likely cluster
members those galaxies in the photometric redshift range of 0.14 to
0.26. This selection leads to a sample of 292 (out of 399) galaxies
which are likely cluster members.

\begin{table}
        \caption[]{Photometric data. Running number for
         galaxies in the present sample, ID (Col.~1); right ascension 
	 and declination (Col.~2, Col.~3) B-, V-,R- and K-band Kron
         magnitude (Col.~4, Col.~5, Col.~6, Col.~7); heliocentric
         corrected redshift $\mathrm{z}$ (Col.~8); photometric redshift
         (Col.~7).}
         \label{phot_cat}
{\footnotesize
           $$ 
           \begin{array}{c c c c c c c c c}
            \hline
            \noalign{\smallskip}
            \hline
            \noalign{\smallskip}

\mathrm{ID} & \mathrm{\alpha} & \mathrm{\delta}  & \mathrm{B} &
            \mathrm{V} & \mathrm{R}  & \mathrm{K} & \mathrm{z_{spec}} &
            \mathrm{z_{phot}}\\
            \hline
            \noalign{\smallskip}   
     1 & 01\ 31\ 47.23 & -13\ 35\ 31.2 &   21.97\pm0.02 &   20.35\pm0.03 &   19.73\pm0.01 &   17.06\pm0.02 & \ \ \ \ \ \ \ \ \ ... \ \ \ \ \ \ \ \ \ & 0.155\\
     2 & 01\ 31\ 49.46 & -13\ 37\ 26.9 &   19.12\pm0.01 &   18.02\pm0.02 &   17.57\pm0.01 &   15.11\pm0.01 & 0.2140\pm0.0002 & 0.170\\
     3 & 01\ 31\ 55.10 & -13\ 37\ 24.1 &   21.92\pm0.01 &   21.27\pm0.03 &   20.63\pm0.01 &   18.71\pm0.09 & \ \ \ \ \ \ \ \ \ ... \ \ \ \ \ \ \ \ \ & 0.451\\
     4 & 01\ 31\ 55.68 & -13\ 37\ 19.9 &   23.09\pm0.03 &   21.43\pm0.03 &   20.94\pm0.01 &   18.18\pm0.07 & \ \ \ \ \ \ \ \ \ ... \ \ \ \ \ \ \ \ \ & 0.147\\
     5 & 01\ 31\ 51.56 & -13\ 37\ 18.0 &   22.03\pm0.02 &   20.46\pm0.02 &   19.67\pm0.01 &   16.54\pm0.01 & \ \ \ \ \ \ \ \ \ ... \ \ \ \ \ \ \ \ \ & 0.209\\
     6 & 01\ 31\ 51.28 & -13\ 37\ 17.5 &   22.54\pm0.02 &   21.06\pm0.02 &   20.34\pm0.01 &   17.47\pm0.03 & \ \ \ \ \ \ \ \ \ ... \ \ \ \ \ \ \ \ \ & 0.160\\
     7 & 01\ 31\ 56.40 & -13\ 37\ 18.2 &   22.17\pm0.02 &   20.65\pm0.02 &   19.97\pm0.01 &   16.93\pm0.02 & 0.2080\pm0.0003 & 0.155\\
     8 & 01\ 31\ 54.59 & -13\ 37\ 04.3 &   22.56\pm0.03 &   19.95\pm0.03 &   20.38\pm0.01 &   17.27\pm0.03 & \ \ \ \ \ \ \ \ \ ... \ \ \ \ \ \ \ \ \ & 0.145\\
     9 & 01\ 31\ 55.12 & -13\ 37\ 04.7 &   20.96\pm0.01 &   18.93\pm0.02 &   18.58\pm0.01 &   15.36\pm0.01 & 0.2118\pm0.0003 & 0.242\\
    10 & 01\ 31\ 55.18 & -13\ 36\ 57.9 &   20.74\pm0.01 &   18.94\pm0.02 &   18.40\pm0.01 &   15.25\pm0.01 & 0.2150\pm0.0002 & 0.186\\
    11 & 01\ 31\ 51.07 & -13\ 36\ 27.6 &   22.96\pm0.03 &   20.49\pm0.03 &   20.78\pm0.01 &   17.91\pm0.05 & \ \ \ \ \ \ \ \ \ ... \ \ \ \ \ \ \ \ \ & 0.145\\
    12 & 01\ 31\ 52.29 & -13\ 36\ 58.3 &   19.53\pm0.01 &   17.37\pm0.02 &   17.19\pm0.01 &   14.14\pm0.00 & 0.2024\pm0.0002 & 0.186\\
    13 & 01\ 31\ 51.79 & -13\ 36\ 54.8 &   22.09\pm0.02 &   20.20\pm0.02 &   19.88\pm0.01 &   16.58\pm0.02 & 0.2074\pm0.0002 & 0.155\\
    14 & 01\ 31\ 52.52 & -13\ 36\ 27.5 &   21.54\pm0.01 &   18.39\pm0.02 &   19.33\pm0.01 &   14.90\pm0.02 & 0.2110\pm0.0003 & 0.209\\
    15 & 01\ 31\ 53.25 & -13\ 36\ 44.2 &   22.78\pm0.03 &   19.50\pm0.02 &   19.92\pm0.01 &   16.06\pm0.06 & 0.2098\pm0.0004 & 0.165\\
    16 & 01\ 31\ 52.87 & -13\ 36\ 35.2 &   21.51\pm0.02 &   19.79\pm0.02 &   19.28\pm0.01 &   16.29\pm0.04 & \ \ \ \ \ \ \ \ \ ... \ \ \ \ \ \ \ \ \ & 0.145\\
    17 & 01\ 31\ 52.57 & -13\ 36\ 44.2 &   21.35\pm0.01 &   19.38\pm0.02 &   18.75\pm0.01 &   15.66\pm0.01 & \ \ \ \ \ \ \ \ \ ... \ \ \ \ \ \ \ \ \ & 0.170\\
    18 & 01\ 31\ 52.53 & -13\ 36\ 40.5 &   18.76\pm0.01 &   17.00\pm0.02 &   16.41\pm0.01 &   12.77\pm0.00 & 0.2097\pm0.0002 & 0.222\\
    19 & 01\ 31\ 50.97 & -13\ 36\ 49.6 &   21.27\pm0.01 &   19.64\pm0.02 &   19.06\pm0.01 &   16.12\pm0.01 & 0.2042\pm0.0003 & 0.155\\
    20 & 01\ 31\ 56.22 & -13\ 36\ 46.8 &   20.16\pm0.01 &   18.43\pm0.02 &   18.10\pm0.01 &   15.19\pm0.01 & 0.2098\pm0.0003 & 0.150\\
    21 & 01\ 31\ 48.64 & -13\ 36\ 46.6 &   22.76\pm0.03 &   21.36\pm0.03 &   20.58\pm0.01 &   17.63\pm0.03 & \ \ \ \ \ \ \ \ \ ... \ \ \ \ \ \ \ \ \ & 0.180\\
    22 & 01\ 31\ 51.32 & -13\ 36\ 56.8 &   19.84\pm0.01 &   18.10\pm0.02 &   17.49\pm0.01 &   14.35\pm0.00 & 0.2068\pm0.0002 & 0.186\\
    23 & 01\ 31\ 57.70 & -13\ 36\ 43.6 &   22.83\pm0.03 &   21.39\pm0.03 &   20.68\pm0.01 &   17.76\pm0.03 & \ \ \ \ \ \ \ \ \ ... \ \ \ \ \ \ \ \ \ & 0.180\\
    24 & 01\ 31\ 56.92 & -13\ 36\ 21.2 &   19.73\pm0.01 &   18.75\pm0.02 &   18.16\pm0.01 &   15.39\pm0.01 & \ \ \ \ \ \ \ \ \ ... \ \ \ \ \ \ \ \ \ & 0.281\\
    25 & 01\ 31\ 57.33 & -13\ 36\ 32.8 &   22.30\pm0.02 &   20.91\pm0.03 &   20.13\pm0.01 &   17.34\pm0.03 & \ \ \ \ \ \ \ \ \ ... \ \ \ \ \ \ \ \ \ & 0.160\\
    26 & 01\ 31\ 48.71 & -13\ 36\ 25.0 &   21.89\pm0.02 &   20.27\pm0.02 &   19.46\pm0.01 &   16.33\pm0.01 & \ \ \ \ \ \ \ \ \ ... \ \ \ \ \ \ \ \ \ & 0.186\\
    27 & 01\ 31\ 54.57 & -13\ 36\ 25.8 &   23.26\pm0.04 &   21.64\pm0.03 &   20.46\pm0.01 &   16.89\pm0.02 & \ \ \ \ \ \ \ \ \ ... \ \ \ \ \ \ \ \ \ & 0.425\\
    28 & 01\ 31\ 51.26 & -13\ 36\ 20.9 &   22.42\pm0.02 &   18.91\pm0.03 &   20.24\pm0.01 &   17.19\pm0.03 & \ \ \ \ \ \ \ \ \ ... \ \ \ \ \ \ \ \ \ & 0.175\\
    29 & 01\ 31\ 53.33 & -13\ 36\ 31.3 &   20.63\pm0.01 &   18.52\pm0.02 &   18.46\pm0.01 &   15.44\pm0.01 & 0.2094\pm0.0002 & 0.186\\
    30 & 01\ 31\ 53.84 & -13\ 36\ 13.0 &   20.50\pm0.01 &   18.90\pm0.02 &   18.11\pm0.01 &   14.94\pm0.00 & 0.2085\pm0.0002 & 0.242\\
    31 & 01\ 31\ 49.36 & -13\ 36\ 06.4 &   22.05\pm0.02 &   20.53\pm0.02 &   19.76\pm0.01 &   16.78\pm0.01 & 0.2133\pm0.0004 & 0.165\\
    32 & 01\ 31\ 49.83 & -13\ 36\ 11.2 &   21.42\pm0.01 &   20.15\pm0.02 &   19.41\pm0.01 &   16.38\pm0.01 & 0.2123\pm0.0003 & 0.215\\
    33 & 01\ 31\ 52.97 & -13\ 36\ 22.3 &   21.82\pm0.02 &   20.16\pm0.02 &   19.40\pm0.01 &   16.33\pm0.01 & \ \ \ \ \ \ \ \ \ ... \ \ \ \ \ \ \ \ \ & 0.186\\
    34 & 01\ 31\ 50.86 & -13\ 36\ 03.8 &   20.41\pm0.01 &   18.78\pm0.02 &   18.03\pm0.01 &   14.85\pm0.00 & 0.2078\pm0.0002 & 0.225\\
    35 & 01\ 31\ 50.31 & -13\ 36\ 01.4 &   23.47\pm0.05 &   21.57\pm0.03 &   20.90\pm0.01 &   17.84\pm0.03 & \ \ \ \ \ \ \ \ \ ... \ \ \ \ \ \ \ \ \ & 0.186\\
    36 & 01\ 31\ 53.67 & -13\ 36\ 03.8 &   23.18\pm0.04 &   21.67\pm0.03 &   20.95\pm0.01 &   18.19\pm0.06 & \ \ \ \ \ \ \ \ \ ... \ \ \ \ \ \ \ \ \ & 0.170\\
    37 & 01\ 31\ 50.36 & -13\ 35\ 52.9 &   22.43\pm0.02 &   20.83\pm0.02 &   20.00\pm0.01 &   17.01\pm0.02 & \ \ \ \ \ \ \ \ \ ... \ \ \ \ \ \ \ \ \ & 0.170\\
    38 & 01\ 31\ 56.88 & -13\ 35\ 51.6 &   22.66\pm0.03 &   21.21\pm0.03 &   20.43\pm0.01 &   17.53\pm0.03 & \ \ \ \ \ \ \ \ \ ... \ \ \ \ \ \ \ \ \ & 0.191\\
    39 & 01\ 31\ 30.91 & -13\ 24\ 42.8 &   23.19\pm0.03 &    \ \ \ \ \ ...\ \ \ \ \ &   20.54\pm0.01 &   17.29\pm0.03 & \ \ \ \ \ \ \ \ \ ... \ \ \ \ \ \ \ \ \ & 0.289\\
    40 & 01\ 31\ 22.10 & -13\ 24\ 37.5 &   21.17\pm0.01 &    \ \ \ \ \ ...\ \ \ \ \ &   18.83\pm0.01 &   15.77\pm0.02 & \ \ \ \ \ \ \ \ \ ... \ \ \ \ \ \ \ \ \ & 0.203\\
    41 & 01\ 31\ 33.44 & -13\ 25\ 02.8 &   21.78\pm0.02 &    \ \ \ \ \ ...\ \ \ \ \ &   19.19\pm0.01 &   15.68\pm0.01 & \ \ \ \ \ \ \ \ \ ... \ \ \ \ \ \ \ \ \ & 0.425\\
    42 & 01\ 31\ 25.47 & -13\ 25\ 15.5 &   23.26\pm0.04 &    \ \ \ \ \ ...\ \ \ \ \ &   20.82\pm0.01 &   17.75\pm0.05 & \ \ \ \ \ \ \ \ \ ... \ \ \ \ \ \ \ \ \ & 0.225\\
    43 & 01\ 31\ 28.92 & -13\ 25\ 19.2 &   22.63\pm0.03 &    \ \ \ \ \ ...\ \ \ \ \ &   20.28\pm0.01 &   17.25\pm0.03 & \ \ \ \ \ \ \ \ \ ... \ \ \ \ \ \ \ \ \ & 0.215\\
    44 & 01\ 31\ 26.99 & -13\ 25\ 07.5 &   22.57\pm0.04 &    \ \ \ \ \ ...\ \ \ \ \ &   19.90\pm0.01 &   16.86\pm0.03 & \ \ \ \ \ \ \ \ \ ... \ \ \ \ \ \ \ \ \ & 0.273\\
    45 & 01\ 31\ 26.50 & -13\ 25\ 13.7 &   20.85\pm0.01 &    \ \ \ \ \ ...\ \ \ \ \ &   18.25\pm0.01 &   14.91\pm0.01 & \ \ \ \ \ \ \ \ \ ... \ \ \ \ \ \ \ \ \ & 0.289\\
    46 & 01\ 31\ 20.43 & -13\ 25\ 20.0 &   21.95\pm0.02 &    \ \ \ \ \ ...\ \ \ \ \ &   19.53\pm0.01 &   15.92\pm0.01 & \ \ \ \ \ \ \ \ \ ... \ \ \ \ \ \ \ \ \ & 0.538\\
    47 & 01\ 31\ 33.65 & -13\ 25\ 32.6 &   20.61\pm0.01 &    \ \ \ \ \ ...\ \ \ \ \ &   18.39\pm0.01 &   15.37\pm0.01 & \ \ \ \ \ \ \ \ \ ... \ \ \ \ \ \ \ \ \ & 0.197\\
    48 & 01\ 31\ 30.07 & -13\ 26\ 04.7 &   23.01\pm0.03 &    \ \ \ \ \ ...\ \ \ \ \ &   20.46\pm0.01 &   17.20\pm0.03 & \ \ \ \ \ \ \ \ \ ... \ \ \ \ \ \ \ \ \ & 0.273\\
    49 & 01\ 31\ 17.43 & -13\ 26\ 04.3 &   22.60\pm0.02 &    \ \ \ \ \ ...\ \ \ \ \ &   20.78\pm0.01 &   17.56\pm0.03 & \ \ \ \ \ \ \ \ \ ... \ \ \ \ \ \ \ \ \ & 0.555\\
    50 & 01\ 31\ 25.21 & -13\ 26\ 33.0 &   22.90\pm0.03 &    \ \ \ \ \ ...\ \ \ \ \ &   20.59\pm0.01 &   17.57\pm0.03 & \ \ \ \ \ \ \ \ \ ... \ \ \ \ \ \ \ \ \ & 0.206\\
    51 & 01\ 31\ 24.65 & -13\ 26\ 08.0 &   20.21\pm0.01 &    \ \ \ \ \ ...\ \ \ \ \ &   17.81\pm0.01 &   14.77\pm0.01 & \ \ \ \ \ \ \ \ \ ... \ \ \ \ \ \ \ \ \ & 0.229\\
    52 & 01\ 31\ 34.11 & -13\ 26\ 35.2 &   22.65\pm0.03 &    \ \ \ \ \ ...\ \ \ \ \ &   20.48\pm0.01 &   16.61\pm0.02 & \ \ \ \ \ \ \ \ \ ... \ \ \ \ \ \ \ \ \ & 0.682\\
    53 & 01\ 31\ 29.12 & -13\ 26\ 38.8 &   23.32\pm0.05 &    \ \ \ \ \ ...\ \ \ \ \ &   20.61\pm0.01 &   16.68\pm0.02 & \ \ \ \ \ \ \ \ \ ... \ \ \ \ \ \ \ \ \ & 0.606\\
    54 & 01\ 31\ 33.01 & -13\ 26\ 46.9 &   21.74\pm0.01 &    \ \ \ \ \ ...\ \ \ \ \ &   19.62\pm0.01 &   16.73\pm0.02 & \ \ \ \ \ \ \ \ \ ... \ \ \ \ \ \ \ \ \ & 0.170\\
    55 & 01\ 32\ 00.96 & -13\ 27\ 06.0 &   21.03\pm0.01 &    \ \ \ \ \ ...\ \ \ \ \ &   18.79\pm0.01 &   15.79\pm0.02 & \ \ \ \ \ \ \ \ \ ... \ \ \ \ \ \ \ \ \ & 0.209\\
    56 & 01\ 31\ 45.29 & -13\ 27\ 14.6 &   20.83\pm0.01 &    \ \ \ \ \ ...\ \ \ \ \ &   18.37\pm0.01 &   15.24\pm0.00 & \ \ \ \ \ \ \ \ \ ... \ \ \ \ \ \ \ \ \ & 0.235\\
    57 & 01\ 31\ 48.70 & -13\ 27\ 28.6 &   21.40\pm0.01 &    \ \ \ \ \ ...\ \ \ \ \ &   19.87\pm0.01 &   16.93\pm0.02 & \ \ \ \ \ \ \ \ \ ... \ \ \ \ \ \ \ \ \ & 0.571\\
    58 & 01\ 31\ 51.61 & -13\ 27\ 34.5 &   22.55\pm0.03 &    \ \ \ \ \ ...\ \ \ \ \ &   20.69\pm0.01 &   17.37\pm0.03 & \ \ \ \ \ \ \ \ \ ... \ \ \ \ \ \ \ \ \ & 0.571\\
    59 & 01\ 31\ 22.90 & -13\ 27\ 32.5 &   22.76\pm0.03 &    \ \ \ \ \ ...\ \ \ \ \ &   20.35\pm0.01 &   17.18\pm0.03 & \ \ \ \ \ \ \ \ \ ... \ \ \ \ \ \ \ \ \ & 0.250\\
    60 & 01\ 31\ 40.99 & -13\ 27\ 22.2 &   20.97\pm0.01 &    \ \ \ \ \ ...\ \ \ \ \ &   18.82\pm0.01 &   16.05\pm0.01 & \ \ \ \ \ \ \ \ \ ... \ \ \ \ \ \ \ \ \ & 0.165\\
            \noalign{\smallskip}	     		    
            \hline			    		    
         \end{array}
     $$ 
}
         \end{table}

\subsection{Catalogues}
\label{sec:33}

The new spectroscopic catalogue is presented in
Table~\ref{spec_catalogue}, which contains the sky coordinates $\alpha$
(Col. 1) and $\delta$ (Col. 2), and the spectroscopic redshift
(Col. 3).

The catalogue of galaxies with reliable photometric redshifts (
$\delta z / 1+z < $ 0.07) is presented in Table~\ref{phot_cat},
available in electronic format, which includes the identification number
of each galaxy, ID (Col. 1), right ascension and declination (J2000),
$\alpha$ and $\delta $ (Cols. 2 and 3), $B$ (Col. 4), $V$ (Col. 5) and
$R$ magnitudes when available (Col. 6), $K$ magnitudes (Col. 7),
heliocentric corrected spectroscopic redshift, z$_{spec}$ (Col. 8), if
available, and the photometric redshift, z$_{phot}$ (Col. 9).

\section{Global cluster properties}
\label{sec:4}

\subsection{NIR luminosity function of cluster galaxies}
\label{sec:41}

\subsubsection{K-band Magnitudes and Completeness}
\label{KBANDMAG}

To perform source detection, we ran SExtractor (Bertin \& Arnouts
\citeyear{ber96}) on the $K$-band mosaic of ABCG\,209, obtaining a
$K$-band catalogue with a total of 2628 objects. For each source, we
measured aperture magnitudes within a $5{\arcsec}$ diameter aperture,
and Kron magnitudes within an aperture of diameter $\alpha \cdot
{r_{\rm K}}$, where ${r_{\rm K}}$ is the Kron radius (Kron 1980).  We
chose $\alpha =2.2$, for which the Kron magnitude is expected to
enclose 92\% of the total flux, and we computed the total magnitude,
$K$, by subtracting 0.08 mag from the Kron magnitudes.

\begin{figure}
\hbox{
{\resizebox{7cm}{!}{\includegraphics{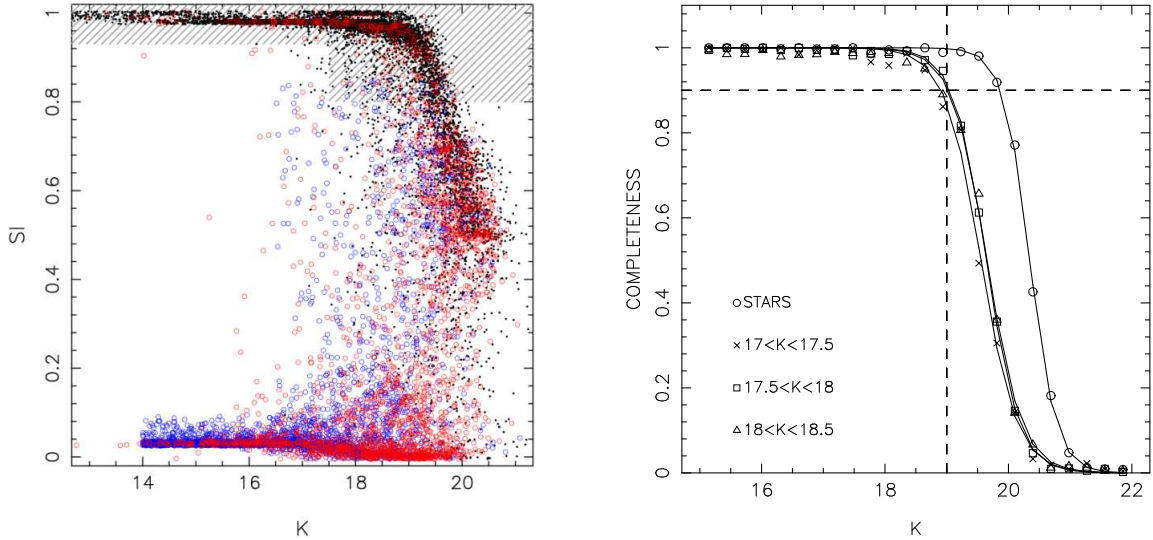}}}
\hfill
\hspace{1cm}
{\resizebox{7cm}{!}{\includegraphics{fig3b.ps}}}
}
\caption{{\it Left panel:} Stellarity index versus $K$-band magnitude
  diagram for all sources detected in the $K$-band mosaic of ABCG\,209
  (red circles), for simulated galaxies (blue circles), and simulated
  stars (black dots). Shaded area indicates the selection criteria
  used to identify stars. {\it Right panel:} $K$-band completeness
  function.  Crosses, squares and triangles denote the completeness
  percentage as estimated by simulating galaxies in different
  magnitude bins (see the text) as shown in the lower left corner of
  the panel.  The completeness percentage of stars is plotted by the
  circles. The solid curves represent the fits of the completeness
  functions by Fermi-Dirac functions. The dashed lines mark a
  completeness level of $90\%$ for the galaxy catalogue.}
\label{completezza}
\end{figure}

As shown in Fig.~\ref{completezza} (left panel), objects were
classified as stars and galaxies according to the distribution of
sources in the $SI$ versus Kron magnitude diagram, where $SI$ is the
stellarity index parameter of SExtractor.  We classified as stars
those objects whose $SI$ value is larger than a given threshold,
$SI_{\rm min}$. The value of $SI_{\rm min}$ was chosen by adding
simulated stars and galaxies to the $K$-band mosaic, and measuring
their $SI$ and ${m_{\rm K}}$ parameters by means of SExtractor, in
the same way as for real sources.  Simulated stars and galaxies were
randomly generated in a magnitude range of $K =14$ mag to $K =21$
mag. In order to create the simulated images, we modelled real sources
in the $K$-band mosaic as the sum of three two-dimensional Moffat
functions.  The Moffat models were scaled according to the magnitude
of simulated objects and added randomly to the $K$-band mosaic.  In
the case of stars, the modelling was performed by fitting non-saturated
sources with $K <16$ mag and $SI>0.9$.  For galaxies, we fitted
objects with $SI<0.2$, in a magnitude range of $K_{\rm min}$ to
$K_{\rm max}$.  We chose $K_{\rm min}=17$ mag and $K_{\rm max}=18.5$
mag, and we also varied the values of $K_{\rm min}$ and $K_{\rm max}$
in order to test the robustness of our completeness estimates (see
below) with respect to these parameters.  Fig.~\ref{completezza} shows
the distribution of simulated objects in the $SI$ versus $K$ diagram.
At magnitudes brighter than $K \sim 17.5$ mag, stars and galaxies are
clearly well separated, with this separation disappearing at fainter
magnitudes.  In fact, for $K \gtrsim 17.5$ mag, simulated stars have a
stellarity index $\gtrsim 0.4$, while almost all simulated galaxies
have $SI \lesssim 0.8$.  In order to minimise the fraction of
misclassified stars and galaxies, we adopted a value of $SI_{\rm min}$
that depends on the source magnitude, with $SI_{\rm min} =0.95$ for $K
\le 17.5$ mag and $SI_{\rm min} = 0.8$ for $K > 17.5$ mag. The
corresponding star/galaxy classification leads to a sample of $2334$
galaxies in the $K$-band field of ABCG\,209.  The completeness of the
$K$-band catalogue was estimated by measuring the percentage of
simulated galaxies and stars which are recovered by SExtractor as a
function of $K$.  The completeness functions are shown in
Fig.~\ref{completezza} (right panel) for both stars and galaxies.  The
figure shows that the $K$-band galaxy catalogue of ABCG\,209 is
complete at more than 90$\%$ down to $K \sim 19$ mag, with the
completeness level falling rapidly to $\sim 50 \%$ at 19.5, and to
$\sim 20\%$ at $K \sim 20$ mag.  As one would expect, the stellar
catalogue is characterised by a higher completeness at a given
magnitude, being more than 90$\%$ complete down to $K \sim 19.8$ mag.
The figure also shows that the galaxy completeness function does not
depend significantly on the magnitude range of the sources used to
create simulated galaxies.  In fact, changing the values of $K_{\rm
min}$ and $K_{\rm max}$, as shown in the figure, we found that the
completeness estimates do not change significantly.

\subsubsection{K-band Luminosity function}
\begin{figure*}
\center
\includegraphics[]{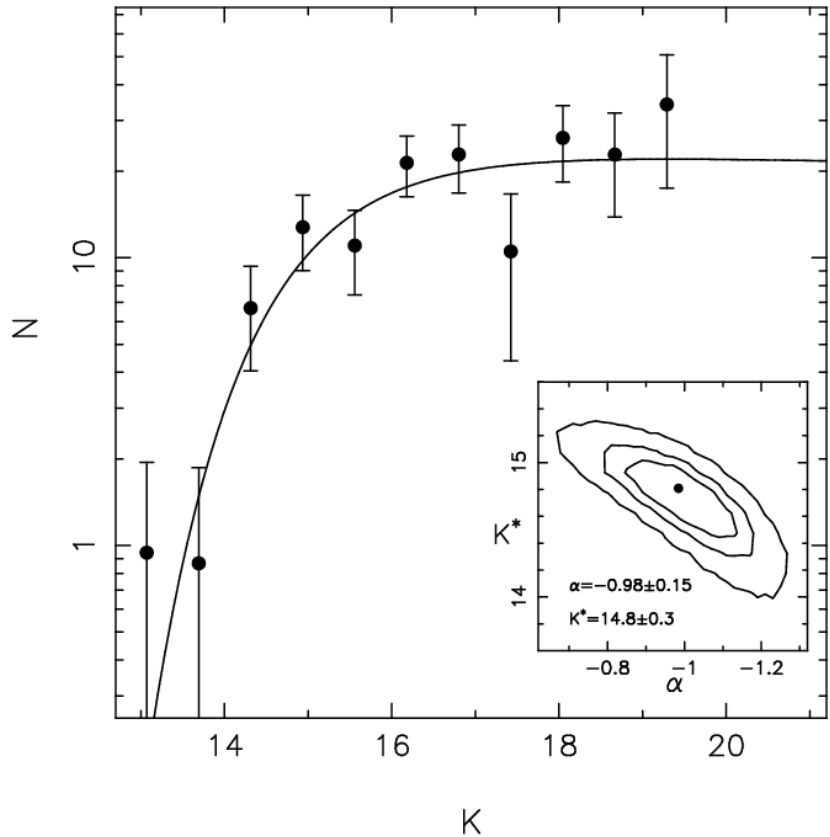}
\caption{$K$-band luminosity function of ABCG\,209.  The solid curve
  shows the best-fitting Schechter function. The smaller panel in the
  lower-right corner of the figure shows the joint probability
  contours of $K^\star$ and $\alpha$ best-fitting values.  Black
  contours correspond to $1\sigma$, $2\sigma$, and $3\sigma$ standard
  probability levels, respectively. The best-fitting value of
  $K^\star$ and $\alpha$ are also reported in the lower-left corner of
  the smaller panel.}
\label{LF}
\end{figure*}

The $K$-band LF of ABCG\,209 was derived in a square region of
4.7 arcmin side size around the cluster center, by computing number
counts of galaxies brighter than $K=19.5$ mag.  Using the simulation
results discussed in Sec.~\ref{KBANDMAG}, we corrected number counts
for (i) incompleteness, dividing by the completeness function shown in
Fig.~\ref{completezza} (right panel), and (ii) contamination from
misclassified stars and galaxies, dividing number counts by the
expected fraction of misclassified galaxies and subtracting the
expected number of misclassified stars in each magnitude bin.  Number
counts were then corrected for background/foreground galaxies, by
using field galaxy counts from the Calar Alto Deep Imaging Survey
(CADIS, see~\citealt{HTK01}), a medium deep $K$-band survey with a
total area of $\rm 0.2~deg^2$ and a completeness magnitude of $\rm
19.75$ mag.  Since our $K$-band photometry is $50\%$ complete down to
$\rm m_{\rm K}=19.5$ mag, the CADIS data are a suitable dataset for
the estimation of field galaxy counts.  The resulting LF of ABCG\,209
\, and its fit with a single Schechter function, are shown in
Fig.~\ref{LF}.  The error bars on number counts take into account
Poissonian uncertainties on both field and cluster counts. The LF fit
was performed by a $\chi^2$ minimisation routine, accounting for the
finite size of magnitude bins by integrating the Schechter function in
each magnitude bin.  Since brightest cluster galaxies usually prevent
a good fit of the LF to be achieved with a single Schechter function
(see e.g.~\citealt{dPE98} and~\citealt{dPS99}), the cD galaxy of
ABCG\,209 was excluded from the fitting. The Schechter function fit,
whose parameters are the faint end slope, $\alpha$, and the
characteristic magnitude, $K^\star$, provides a good description of
cluster galaxy counts.  The only magnitude bin showing a larger
deviation is that at $K=17.5$ mag, where a marginal ($\sim 1.5 \sigma$
significant) deficiency of cluster galaxies is found. Using an $R-K$
colour of $\sim 3.1$, which is typical of early-type cluster galaxies
at $z \sim 0.2$ (e.g.~\cite{lab04}, hereafter LM04), the magnitude of
$K=17.5$ mag can be transformed to $R \sim 20.6$ mag, that corresponds
to the magnitude bin ($20<R<21$ mag) where Mercurio et
al. (\citeyear{mer03b}) find some indication of a dip in the optical
luminosity functions of ABCG\,209.  The best-fitting value of the
$K$-band Schechter function parameters turn out to be
$\alpha=-0.98 \pm 0.15$ and $\rm K^* = 14.80 \pm 0.30$ mag,
respectively.  The uncertainties on $\alpha$ and $K^\star$ were
derived by randomly shifting galaxy number counts according to their
uncertainties, and then re-computing the best-fitting Schechter
function.  Joint probability contours for $\alpha$ and $K^\star$ are
shown in Fig.~\ref{completezza}.  The above value of $\alpha$ is fully
consistent with those found from previous studies of NIR luminosity
functions of cluster galaxies, such as $\alpha= \sim
1.0$~\citep{dPE98}, $\alpha \sim -1.3$~\citep{AnP00}, and $\alpha \sim
-1.2$ (LM04).

\subsection{Dynamical properties}
\label{sec:42}

In a previous paper we investigated the dynamical status of the galaxy
cluster ABCG\,209, analysing a sample of 119 spectra with 8 \AA \ \
resolution and selecting 112 galaxies belonging to the cluster
(Mercurio et al. \citeyear{mer03a}). Hereafter we will refer to this
sample as the ``old sample''. Among the old sample, 22 galaxies were
observed again at higher resolution within the two masks together with
29 new galaxies. Combining the old and new redshifts we obtained a
total sample of 148 galaxies. In Table \ref{spec_catalogue} we report
spectroscopic redshift measurements of the new observed galaxies.

\subsubsection{Member selection}

\begin{figure*}
\center
\includegraphics[width=1.0\textwidth,bb=12 31 425 312,clip]{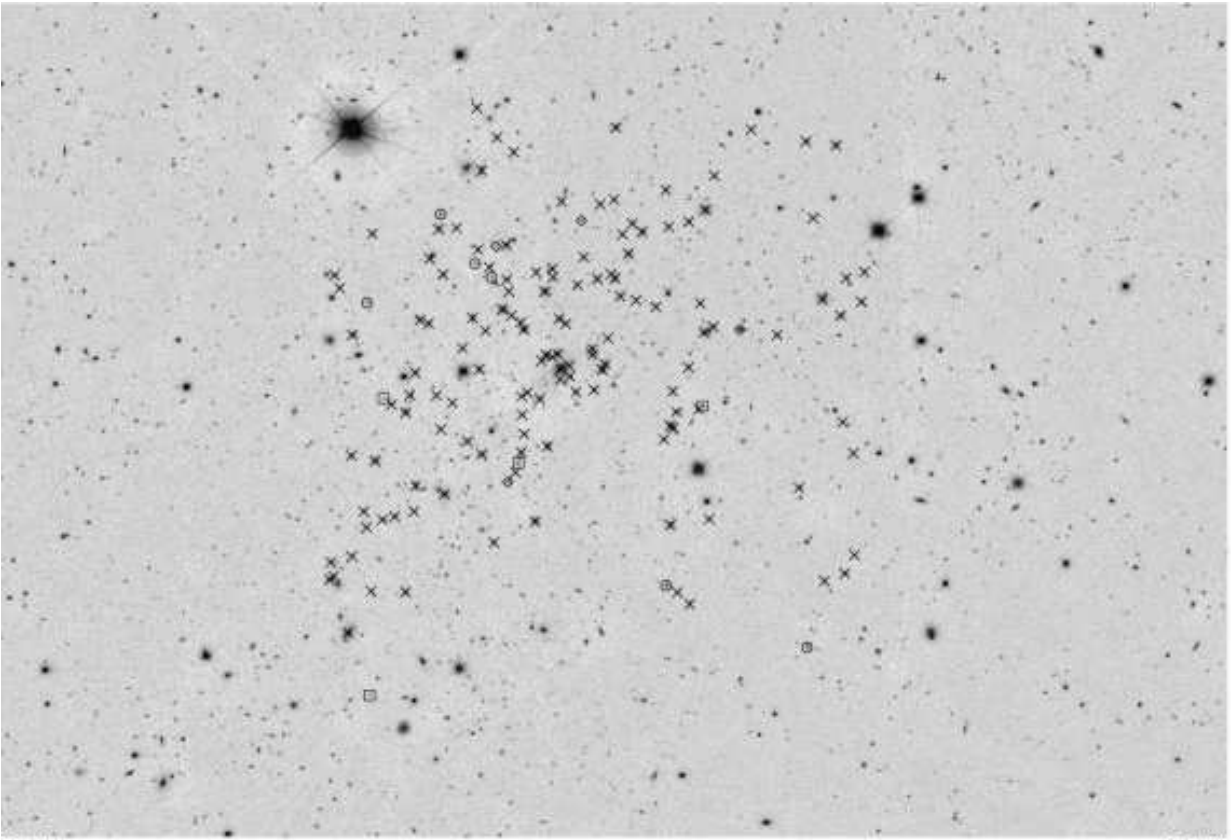}
\caption{CFHT R--band image of ABCG\,209 (North at
top and East to left). Galaxies belonging to the cluster are marked
with crosses, galaxies at ${z\sim0.257}$ with circles, galaxies at
${z\sim0.392}$ with squares and the remaining isolated galaxies with
diamonds.}
\label{fig1_image}
\end{figure*}

\begin{figure*}
\centering
\includegraphics[width=0.5\textwidth]{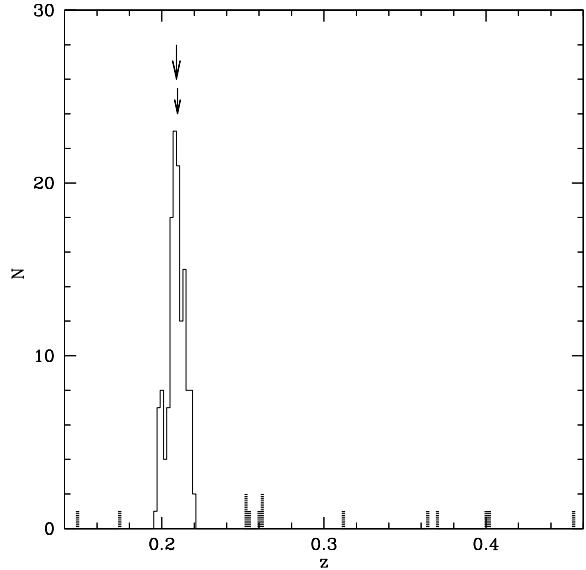}
\caption{Distribution of redshifts for the 148 observed galaxies. The
big and small arrows indicate the mean cluster redshift and
the redshift of the cD galaxy, respectively. Black areas mark
galaxies not belonging to the cluster.}
\label{fig2_histo}
\end{figure*}

In order to select cluster members, we analysed the velocity
distribution by applying the one-dimensional adaptive kernel technique
(\citealt{pis93}, as implemented by \citealt{fad96} and
\citealt{gir96}). This procedure confirms the existence of a single
peak at ${z\sim 0.209}$, consisting of 134 cluster members.  Moreover,
this procedure indicates two other possible structures at higher
redshifts, one at ${z\sim0.257}$ with 6 candidate members, and another
at ${z\sim0.392}$ with 4 candidate members. The remaining 4 galaxies
are considered field galaxies. Although the structure at
${z\sim0.257}$ has a high significance (96.6\%), the low number of
galaxies prevents us from assessing the existence of a background
cluster. In Fig.~\ref{fig1_image} we show the 134 cluster members as
crosses, the 6 galaxies at ${z\sim0.257}$ as circles and the 4
galaxies at ${z\sim0.392}$ as squares. Figure~\ref{fig2_histo} shows
the redshift distribution of the 148 observed galaxies.  The mean
redshift of the cluster as derived by the biweight estimator (Beers et
al. \citeyear{bee90}) is ${<z>=0.2090\pm0.0004}$.

We further explored the possibility of contamination by interlopers by
applying the ``shifting gapper'' algorithm of Fadda et
al. (\citeyear{fad96}). This procedure rejects galaxies that are too
far in velocity-space from the main body of cluster galaxies within a
fixed radial bin, shifting along the distance from the cluster
centre. Following the prescriptions of Fadda et
al. (\citeyear{fad96}) we used a gap of 1000 km s$^{-1}\;$ and a bin
of 0.6 h$_{70}^{-1}$ Mpc, or large enough to include at least 15
galaxies.  In this case eight galaxies were rejected (cf. crosses in
Fig. \ref{fig3}).

In order to determine the cluster centre, we applied the
two-dimensional adaptive kernel technique to the galaxy positions. The
centre of the most significant peak ( $\alpha$= 01 31 52.59, $\delta$=
-13 36 45.9) is coincident with the position of the cD galaxy and only
4${\arcsec}$ distant from that of the old sample.

\subsubsection{Velocity dispersion}
\label{sec:422}

We estimated the line-of-sight velocity dispersion (LOSVD),
${\rm\sigma _v}$, using the biweight estimator ({\sc rostat} package;
Beers et al. \citeyear{bee90}). After applying the relativistic
correction and the usual correction for velocity errors (Danese et
al. \citeyear{dan80}), considering only the 126 galaxies selected by
the "shifting gapper'' method, we obtained
${\rm\sigma_v=1268^{+93}_{-84}}$ km s$^{-1}$, where errors were
estimated with the bootstrap method. If we consider all the 134
galaxies we obtain ${\rm\sigma_v=1390^{+91}_{-91}}$ km s$^{-1}$. This
value is slightly larger, but fully consistent with the value computed
excluding the galaxies identified as outliers by the ``shifting
gapper'' method.

\begin{figure*}
\centering
\includegraphics[angle=-90,width=0.5\textwidth]{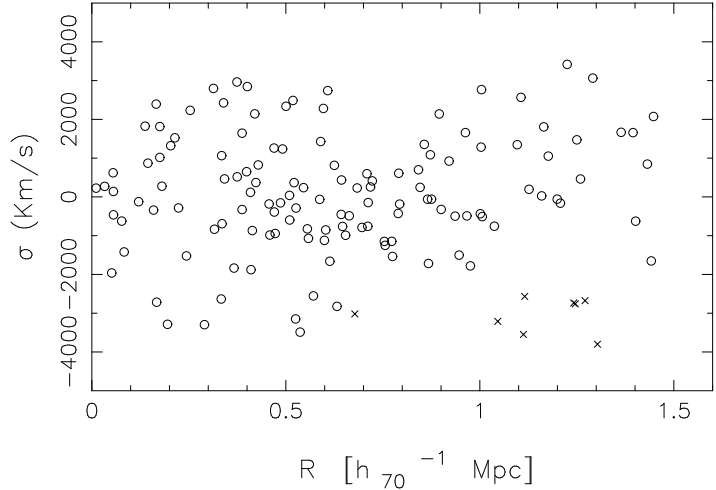}
\caption{Velocity in the cluster rest frame vs. (projected)
clustercentric distance for the 134 selected members. The application
of the ``shifting gapper'' method with a bin of 0.6
h$_{70}^{-1}$ Mpc rejects the  eight galaxies indicated by crosses.}
\label{fig3}
\end{figure*}

\begin{figure*}
\centering
\includegraphics[width=0.8\textwidth]{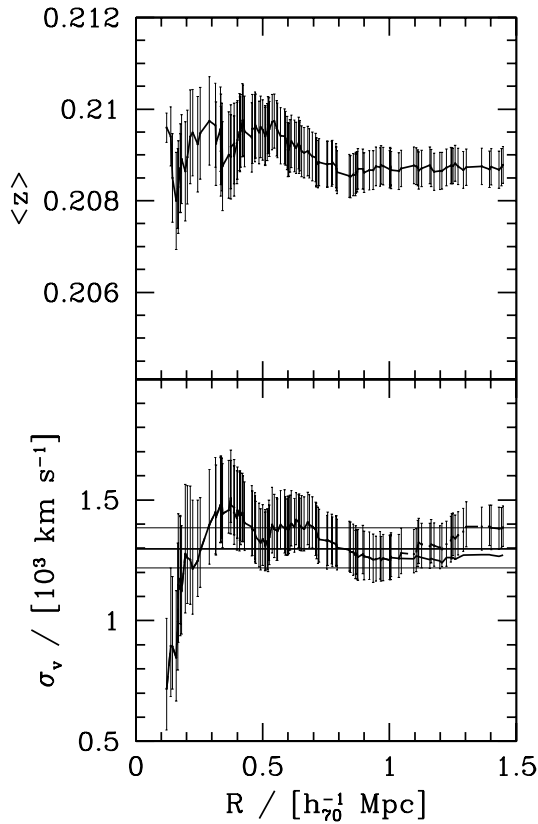}
\caption{Integrated mean redshift and LOSVD profiles
for the 134 galaxies (upper and lower panel, respectively) at a given
projected distance from the cluster centre, estimated by considering
all the galaxies within this radius. The error bars at the $68\%$
c.l. are shown. In the lower panel, solid and dot-dashed lines give the
profile after and before the rejection of eight possible interlopers
according to the ``shifting gapper'' method, respectively. The
horizontal lines represent X--ray temperature with the respective
errors transformed in $\sigma_v$ imposing $\beta_{spec}=1$, where
$\beta_{spec}=\sigma_v^2/(kT/\mu m_p$), with $\mu$ the mean molecular
weight and $m_p$ the proton mass.}
\label{fig4prof}
\end{figure*}

Velocity anisotropies in the galaxy distribution can create problems
in determining the LOSVD. For this reason we plot in
Fig. \ref{fig4prof} the cluster velocity (upper panel) and the
velocity dispersion (lower panel) vs. projected clustercentric
distance. This shows that the ${\rm\sigma_v}$, computed by including
larger and larger regions, decreases in the central cluster regions,
but flattens out in the external regions, suggesting that the value of
the LOSVD is no longer affected by velocity anisotropies. The
rejection of the eight galaxies due to the application of the shifting
gapper (cf. crosses in Fig. \ref{fig3}) leads only to a small
variation in the estimate of ${\rm\sigma _v}$ and in the velocity
dispersion profile (see dot-dashed line in Fig. \ref{fig4prof}) so we
confirm our first result of Mercurio et al. (\citeyear{mer03a}) that
the high value of velocity dispersion is connected to the strong
dynamical evolution of the cluster. In fact, this large value is
already reached at a radius of 0.4-0.5 h$^{-1}_{70}$Mpc, where the
contamination of interlopers is expected to be negligible.

\subsubsection{Velocity dispersion of red sequence galaxies}

In order to further investigate the possible enhancement of the
velocity dispersion due to projection effects from infalling galaxies in
the centre of the cluster we obtain the velocity dispersion profile
considering only galaxies lying on the Colour-Magnitude relation
(CMR). 

We obtained the CMR by fitting the photometric data of the
spectroscopically confirmed cluster members and those galaxies with
photometric redshift in the range $0.14-0.26$ (see Sect. 3.2) with a
biweight algorithm (Beers et al. 1990), obtaining:

\begin{equation}
(R-K)_{\rm CM} = 4.673  - 0.096 \cdot R  \ . 
\label{eqCM}
\end{equation} 

By using Eq.~\ref{eqCM}, we defined as sequence galaxies the sources lying
in the region inside the curves:

\begin{displaymath}
{(R-K)_{\rm seq} = (R-K)_{\rm CM} \pm 2  \cdot \left(\sqrt{\sigma_R^2
    +\sigma_K^2} + 0.05\right) \ , } 
\label{regions}
\end{displaymath} 

\noindent
where we take into account the uncertainty at 2$\sigma$ both on the
$R$ (${\sigma_R}$) and on the $K$ magnitude ( ${\sigma_K}$) as well as
the intrinsic dispersion of the CMR (e.g., Merluzzi et al. \citeyear{mer02},
\citeyear{merl03}).

\begin{figure*}
\center
\includegraphics[width=0.6\textwidth,angle=-90]{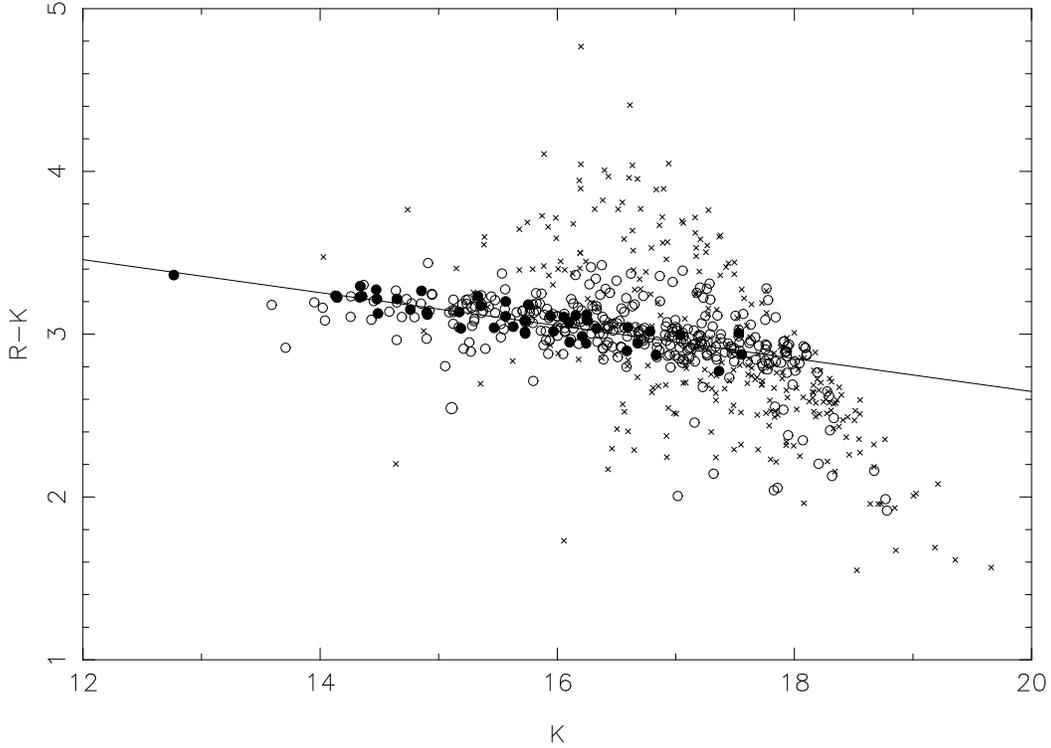}
\caption{The R-K/R colour-magnitude diagram of the cluster galaxy
population.  Open circles indicate galaxies with photometric redshift in
the range 0.15-0.30 and crosses represent the other galaxies with
photometric data.  The solid line indicates the best-fitting C-M
relation of Eq.~\ref{eqCM}, and spectroscopically confirmed member
galaxies identified as belonging to the red sequence through
Eq.~\ref{regions} are indicated by solid circles.}
\label{figCM}
\end{figure*}

\begin{figure*}
\center
\vspace{-5.0cm}
\includegraphics[width=0.8\textwidth]{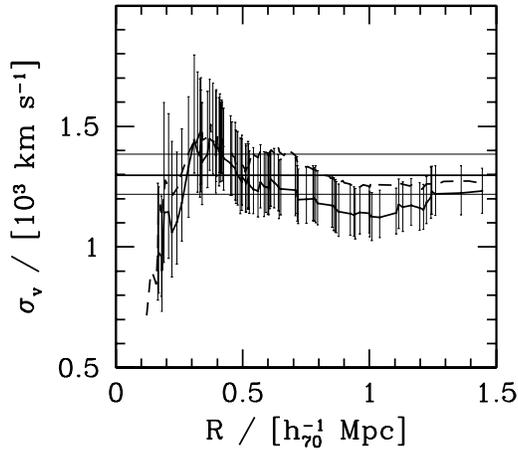}
\caption{Integrated LOS velocity dispersion profile
for the 96 galaxies at a given
projected distance from the cluster centre, estimated by considering
all the galaxies within this radius. The error bars at the $68\%$
c.l. are shown. Solid and dashed lines give the
profile considering only red sequence galaxies and all the cluster
member galaxies after the rejection of eight possible interlopers
according to the ``shifting gapper'' method, respectively. The
horizontal lines represent X--ray temperature with the respective
errors transformed in $\sigma_v$ imposing $\beta_{spec}=1$, where
$\beta_{spec}=\sigma_v^2/(kT/\mu m_p$), with $\mu$ the mean molecular
weight and $m_p$ the proton mass.}
\label{CMprof}
\end{figure*}

We select 96 spectroscopically confirmed member galaxies lying on the CMR
(filled circles in Fig.~\ref{figCM}), obtaining a LOSVD for the red
sequence galaxies of ${\rm\sigma_{v,CMR}=1232^{+94}_{-92}}$ km
s$^{-1}$. By selecting galaxies on the CMR, we expect to include only
early-type galaxies and to minimise the effect of infalling late type
galaxies on the line-of-sight velocity dispersion. However the
measured value of the LOSVD and the velocity profile shown in
Fig.~\ref{CMprof} are fully consistent with those obtained considering
all the sample (see dashed line in Fig.~\ref{CMprof}).

\subsubsection{Virial radius and mass}

Assuming that galaxies are in dynamical equilibrium with the cluster
potential we can determine the cluster mass from the knowledge of
positions and velocities of the member galaxies by applying the virial
theorem. Since the application of the virial theorem is meaningful
only when the system is in dynamical equilibrium within the considered
region, a natural choice is to compute the cluster mass, $M_{\rm
vir}$, inside the radius of virialization, $R_{\rm vir}$, within which
the cluster can be considered to be not far from dynamical
equilibrium.  

Following the prescriptions of Eq.~1 of Girardi \& Mezzetti
(\citeyear{gir01}): $R_{vir}=0.17\times {\rm\sigma_v}/H(z)$, where
$H(z)=H_0(1+z)^{3/2}$ (see also Eq.~8 of Carlberg et
al. \citeyear{car97} for $R_{200}$) and $\rm\sigma_v$ is a ``robust''
estimate of the line-of-sight velocity dispersion. We indicate as
robust a value that is not affected by the presence of velocity
anisotropies in the galaxy orbits that can strongly influence the
central cluster regions. Following Fadda et al. \citeyear{fad96}), we
check the robustness of $\rm\sigma_v$ from the analysis of velocity
dispersion profile. Since the profile flattens out in the external
regions (see Fig.~\ref{fig4prof}), our estimate of the velocity
dispersion is robust (see Sect.~\ref{sec:422}).

In a $\Lambda$CDM model, with $\Omega_M=0.3$, $\Omega_{\Lambda}$=0.7
and $H_0=70$ km s$^{-1}$, the value of
${\rm\sigma_v=1268^{+93}_{-84}}$ km s$^{-1}$ leads to a value of the
virial radius ${R_{\rm vir}\sim 3.42} \ \ h_{70}^{-1}$ Mpc.
In our previous work (Mercurio et al. \citeyear{mer03a}), we
considered a flat universe with $\Omega_M$=1. In this case the
collapsed region for a spherical model corresponds to the region
inside $R_{178}$. For a comparison with our previous estimates we
derive ${R_{178}\sim 2.31}$ $h_{70}^{-1}$ Mpc.

On the hypothesis that the galaxy distribution traces the mass
distribution we obtain from the virial theorem (Limber \& Mathews
\citeyear{lim60}):

\begin{equation}
M_{vir} = \frac{2<v^2>}{G<r_{ij}^{-1}>} = 
\frac{{\rm\sigma}^2 R_{vir}}{G}  \ ,\
\label{mass1}
\end{equation}

\noindent
where $v$ is the galaxy velocity referred to the cluster mean velocity,
$r_{ij}$ is the distance between any pair of galaxies,  $\sigma$ is
the global velocity dispersion.

Assuming that the cluster is spherical, non-rotating system, we can
use the projected velocity dispersion  $\sigma_v$ and the projected
virial radius, $R_{PV}$. Therefore, Eq.~\ref{mass1} becomes:

\begin{equation}
M_{vir} = \frac{3\pi}{2}\frac{{\rm\sigma_P}^2 R_{PV}}{G}  \ .\
\end{equation}

In this formula $R_{PV}$ is the projected radius, equal to twice the
(projected) harmonic radius:
\begin{equation}
R_{PV}=\frac{N(N-1)}{\Sigma_{i>j} R_{ij}^{-1}} \ \ , \ 
\end{equation}
\noindent 
where $R_{ij}^{-1}$ is the projected distance between two galaxies and
$N$ is the number of observed galaxies. The value of $R_{PV}$ depends
on size of the sampled region, and on the quality of the sampling. In
particular, it increases with the cluster aperture $A$ within which it
is estimated.

In our data the cluster is sampled out to a radius of $R \sim 1.5
h_{70}^{-1}$ Mpc or $\sim 0.6 R_{178}$, but we uniformly sample the
region $R_{max} \sim 1.0 h_{70}^{-1}$ Mpc or $\sim 0.4 R_{178}$.  If
we consider only the region of the cluster that is well sampled, we
obtain $R_{PV}=(0.73\pm0.05) h_{70}^{-1}$ Mpc (the error is obtained
through the jackknife method) within $A=R_{max}$.  Although the
cluster is not sampled out to $R_{178}$, we can estimate $R_{PV}$ at
$A=R_{178}$ theoretically from the knowledge of the parameters
$\alpha$ and $R_C$ of the King-like distribution. We can use Eq. 13 of
Girardi et al. (\citeyear{gir98}). In this case $R_{PV}= 1.48
h_{70}^{-1}$ Mpc.

When the system is not entirely included in the observational sample,
the usual form of the virial theorem 2T + V = 0 should be replaced by
2T + V = 3PV, where 3PV is the surface pressure term (e.g., The \& White
\citeyear{the86}). Therefore, a surface pressure term C should be
applied to the mass estimate: 

\begin{equation}
M_{c}=M_{vir} - C \ \ \ . \ 
\end{equation}

The value of the surface pressure term correction $C$ can be obtained
by analysing the velocity dispersion profile, but this procedure
requires several hundreds of member galaxies. Combining data of many
clusters, Girardi \& Mezzetti (\citeyear{gir01}) obtained that
velocities are isotropic and that the correction at $R \sim R_{vir}$
is $C=0.2 M_V$ (e.g. Carlberg et al. \citeyear{car97}, Girardi et
al. \citeyear{gir98}).  Then, following this approach we can estimate
the mass of the system inside the collapsed region, assuming a value
of 20\% for the surface pressure term correction.  This leads to a
value of total mass
${M_{c}(<\!R_{178})=2.12^{+0.47}_{-0.48}\times10^{15}}$ ${h^{-1}_{70}
\ M_{\odot}}$, while in a $\Lambda$CDM model, with the adopted
cosmology ${M_c(<\!R_{\rm vir})=2.95^{+0.80}_{-0.78}\times10^{15}}$
${h^{-1}_{70} \ M_{\odot}}$.

\section{Dynamical analysis}

\subsection{Velocity distribution}

In order to perform a robust description of the velocity distribution
and to detect possible subclumps we applied the 1D KMM algorithm
(Ashman et al. \citeyear{ash94}) to the 126 galaxy sample, selected by
the ``shifting gapper'' method, and to the
original sample of the 134 galaxies. The KMM algorithm fits a
user-specified number of Gaussian distributions to a dataset and
assesses the improvement of that fit over a single Gaussian. In
addition, it provides the maximum-likelihood estimate of the unknown
n-mode Gaussians and an assignment of objects into groups. The KMM
algorithm is most appropriate in situations where theoretical and/or
empirical arguments indicate that a Gaussian model is valid, as in the
case of the cluster velocity distributions, where gravitational
interactions drive the system toward a relaxed configuration with a
Gaussian velocity distribution, but one of the major uncertainties of
this method is the optimal choice of the number of groups for the
partition.

\begin{figure*}
\hspace{1cm}
\includegraphics[width=10cm]{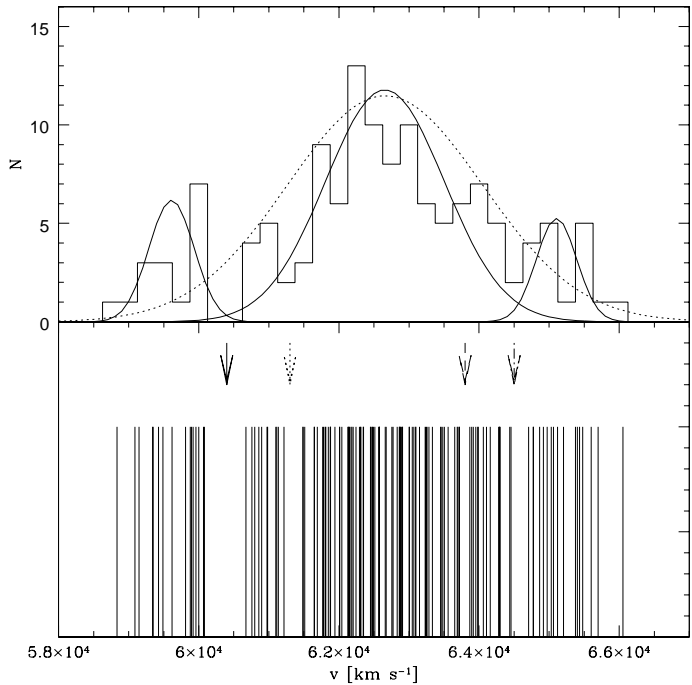}
\caption{Distribution of radial velocities for member galaxies. {\it
Upper panel}: velocity histogram with a bin of 250 km s$^{-1}$, with
the three solid Gaussians showing the best separation into three clumps
as identified with the one--dimensional KMM test (see
text), and the dotted Gaussian indicating the best-fit single Gaussian
function to the velocity distribution.  {\it Lower panel}: stripe density plot. The arrows indicate
the positions of the three significant gaps in the velocity
distribution of member galaxies (see Beers et al. \citeyear{bee91}).}
\label{figgauss}
\end{figure*}

For this reason, we first investigate the presence of gaps in the
velocity distribution using the {\sc rostat} package (Beers et al.
\citeyear{bee90}), which has the advantage of not requiring any {\it a
priori} assumption for the shape of the distribution. Within this
package, a gap is defined as the difference between two contiguous
velocities, weighted by the distances of these velocities with respect to
their average. We considered only normalised gaps larger than 2.54
since, by considering a random sampling of a Gaussian distribution,
such gaps appear only in less than 1.4\% of the cases, independently
of the sample size. We found four significant gaps indicated by
continuous (3.88, 0.05\% of the cases), dotted (3.25, 0.10\% of the
cases), dashed (2.59, 1.4\% of the cases) and dot-dashed (2.57, 1.4\%
of the cases) arrows in the lower panel of Fig. \ref{figgauss}. We
underline that those gaps are found by using all the 134 galaxies in
the sample, but identical results are obtained considering only 126
galaxies.

Using the results of the weighted-gap analysis we divide galaxies into
into five groups and we apply the 1D KMM. From the maximum likelihood
statistics, however, we found that a mixture of five Gaussians do not
give a good description of the data (45.7\% c.l.).  Then we divide
galaxies in four groups and we apply the 1D KMM, but from the maximum
likelihood statistics, we found that the data can be described as a
mixture of four Gaussians at only 89.7\% c.l. On the other hand, from
the maximum likelihood statistics we found that a mixture of three
Gaussians is the best description of the velocity distribution (at
$98.4\%$ c.l.). In the upper panel of Fig. \ref{figgauss} we plot the
velocity distribution of galaxies with, superimposed, the three
Gaussians corresponding to the three identified clumps. The most
prominent gap detected by the {\sc rostat} package is between the
clumps 1 and 2 identified by the KMM algorithm.  Using the one
dimensional KMM algorithm we assigned the member galaxies to individual
groups (n$_1$=16, n$_2$=102, and n$_3$=16 members at mean redshift
z$_1$ = 0.1988, z$_2$ = 0.2090, and z$_3$ = 0.2172) and we estimate
the velocity dispersion of ${\rm\sigma_{v1} = 323^{+47}_{-58}}$ \kss,
${\rm\sigma_{v2} = 847^{+52}_{-49}}$ \kss, and ${\rm\sigma_{v3}
=289^{+57}_{-54} }$ \kss (see Fig.\ref{figgauss}) for each clump. We
note that the old sample of 112 member galaxies did not allow us to
associate galaxies to the less populated substructures and then to
obtain a reliable estimation of their velocity dispersions.

\begin{figure*}
\hspace{1cm}
\includegraphics[bb= 9 347 527 707,clip,width=15cm]{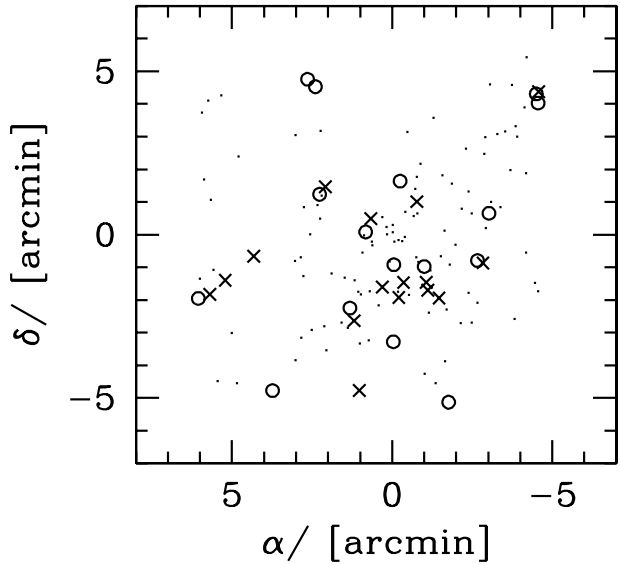}
\vspace{-1.5cm}
\caption
{Distribution of member galaxies separated into three clumps according
the one--dimensional KMM test.  The plot is centred on the cluster
centre. Open circles, dots and crosses indicate clumps 1, 2 and 3, 
having low, intermediate and high velocities, respectively.}
\label{figkmm}
\end{figure*}

Figure \ref{figkmm} shows the spatial distributions of the three
clumps. Differently from the results of our first paper, we find no
evidence for spatial segregation for the three clumps, according to
the two-dimensional Kolmogorov-Smirnov test (hereafter 2DKS-test;
cf. Fasano \& Franceschini \citeyear{fas87}, as implemented by Press
et al. \citeyear{pre92}).

\subsection{3D substructures}

\begin{figure*}
\hspace{0.5cm}
\vspace{-3.5cm}
\includegraphics[width=12cm]{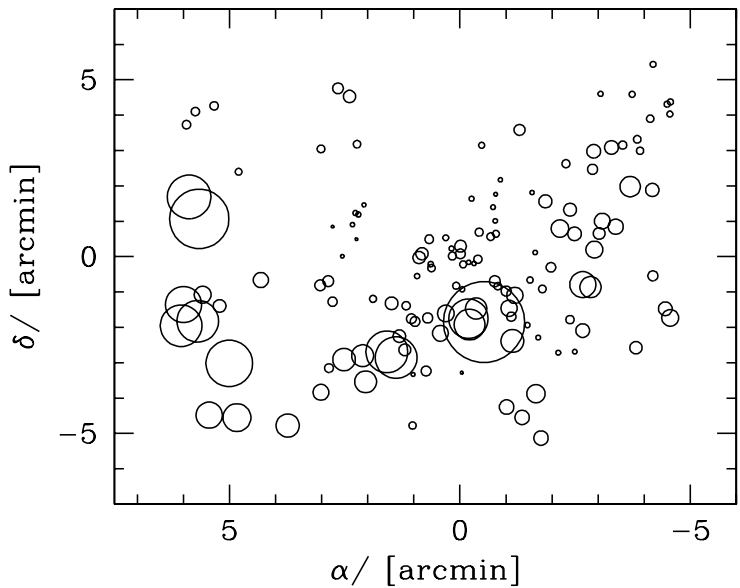}
\caption
{Spatial distribution on the sky of the 134 cluster members, each marked
by a circle: the larger the circle, the larger is the deviation
$\delta$ of the local parameters from the global cluster
parameters. The figure shows evidence for substructure according to
the Dressler \& Schectman test.
The plot is centred on the cluster centre. 
}
\label{figds}
\end{figure*}

In order to check for the presence of substructure, we combined
velocity and position information by computing the $\Delta$-statistic
devised by Dressler \& Schectman (\citeyear{dre88}). This test is
sensitive to spatially compact subsystems having an average velocity
and/or a velocity dispersion different from the global mean
quantities.  We found a value of 198 for the $\Delta$ parameter, which
gives the cumulative deviation of the local kinematical parameters
(velocity mean and velocity dispersion) from the global cluster
parameters. The significance of substructure was checked by running
1000 Monte Carlo simulations, randomly shuffling the galaxy
velocities, obtaining a significance level of $99.8\%$.  This
indicates that the cluster has a complex structure. In
Fig. \ref{figds} we plot the member galaxies marked by circles whose
diameter is proportional to the deviation $\delta $ of the individual
parameters (position and velocity) from the mean cluster parameters.
We confirm with this larger sample the results of the previous
paper. In particular, a group of galaxies with high velocity in the
external East cluster region and another group near the cluster centre
could cause the large values of $\delta $.  We also underline that we
obtain a value of 199 for the $\Delta$ parameter with a significance
level of $99.9\%$, by considering only the 126 galaxies selected with
the shifting gapper method.

Since the technique by Dressler \& Schectman does not allow a direct
identification of galaxies belonging to the detected substructure, we
apply the three dimensional version of the KMM test using
simultaneously galaxy positions and velocities. We use the galaxy
separation obtained in the 1D analysis as first guess of the 3D
analysis.  The algorithm fits a three-group partition at 98.6\% c.l.,
with a galaxy distribution similar to those obtained by the 1D
analysis (n$_1$=16, n$_2$=108, and n$_3$=10 members). On the other
hand the 3D KMM fits a four-group partition at 100.0\% c.l., with a
partition similar to those obtained in the gap analysis performed with
the {\sc rostat} package (n$_1$=16, n$_2$=32, n$_3$=29, and n$_4$=47
members). However, this test seems to be too sensitive to changes in
the initial conditions, in fact with a particular choice of the
initial group partition it is possible to obtain a separation into two
groups significant at 98.0\% c.l.

\section{Structural properties of early-type galaxies}
\label{sec:6}

\subsection{Structural parameters}
\label{sec:61}

We derived the structural parameters, namely half-light radius, $r_e$,
average surface brightness inside $r_e$, $\mu_e$, and Sersic shape
parameter, $n$, in the R and K band for a large sample of
galaxies belonging to ABCG\,209.

Structural parameters were derived for galaxies in the R-, and K-band,
by using the 2DPHOT package (La Barbera et al. \citeyear{lab08}).

We considered the sample of spectroscopically confirmed cluster
members, complemented with the galaxies having photometric redshift in
the range of $0.14-0.26$ (see Sect. \ref{sec:32}).  We found that
for Structural parameters were derived for galaxies with $R<21.0$ mag
and $K<17.5$, i.e. the limits of spectroscopic observations, leading
to a total sample of 327 galaxies. Numerical simulation were performed
in order to check the reliability of the derived structural
parameters, finding that the typical errors were $\delta \mathrm{(log
r_e)} \sim 0.14$ and $\delta \mathrm{(<\mu_e>)} \sim 0.55$ on the
effective radius and surface brightness respectively.

Structural parameters were derived by fitting galaxy images with
seeing-convolved Sersic models.  The point-spread functions (PSFs)
were derived locally from the stars in the R and K images and were
modelled by a sum of 2D Moffat functions, also taking into account
deviations of stellar isophotes from circular symmetry.

Details on the different aspects of the derivation of structural
parameters may be found in La Barbera et al. (\citeyear{lab02}),
(\citeyear{lab03a}), (\citeyear{lab03b}) and (\citeyear{lab08}).

The structural parameters are given in Table~\ref{str_cat} for the
whole sample of 327 galaxies with photometric and spectroscopic
redshifts. We remark that this constitutes so far the largest sample
of (optical plus NIR) structural parameters of galaxies belonging to
one cluster.

\begin{table}
        \caption[]{Structural parameters of galaxies in
         ABCG\,209, according to the photometric redshifts. Right ascension and declination (Col.~1, Col.~2)
         effective radius, mean surface brightness and Sersic index for
         R (Col.~3, Col.~4, Col.~5) and K band (Col.~6, Col.~7, Col.~98)}
         \label{str_cat}
{\footnotesize
           $$ 
           \begin{array}{c c c c c c c c}
            \hline
            \noalign{\smallskip}
            \hline
            \noalign{\smallskip}

            \mathrm{\alpha} & \mathrm{\delta}  & <\mu_R>_e & r_{e,R} &
             \mathrm{n_R}  & <\mu_K>_e & r_{e,K}  &
            \mathrm{n_K}\\
            \hline
            \noalign{\smallskip}   

 01\ 31\ 47.233 & -13\ 35\ 31.20 &  20.6688 &   0.5633 &   4.2156 &  17.1582 &   0.4069 &   2.1719 \\
 01\ 31\ 49.460 & -13\ 37\ 26.92 &  19.8916 &   1.1842 &   1.9660 &  17.5605 &   1.5770 &   6.7775 \\
 01\ 31\ 55.678 & -13\ 37\ 19.92 &  20.6329 &   0.3558 &   3.6644 &  15.7226 &   0.2460 &   1.6739 \\
 01\ 31\ 51.557 & -13\ 37\ 18.00 & \ \ \ \ \ ...\ \ \ \ \ & \ \ \ \ \ ...\ \ \ \ \ & \ \ \ \ \ ...\ \ \ \ \ & \ \ \ \ \ ...\ \ \ \ \ & \ \ \ \ \ ...\ \ \ \ \ & \ \ \ \ \ ...\ \ \ \ \ \\
 01\ 31\ 51.277 & -13\ 37\ 17.55 & \ \ \ \ \ ...\ \ \ \ \ & \ \ \ \ \ ...\ \ \ \ \ & \ \ \ \ \ ...\ \ \ \ \ & \ \ \ \ \ ...\ \ \ \ \ & \ \ \ \ \ ...\ \ \ \ \ & \ \ \ \ \ ...\ \ \ \ \ \\
 01\ 31\ 56.403 & -13\ 37\ 18.23 & \ \ \ \ \ ...\ \ \ \ \ & \ \ \ \ \ ...\ \ \ \ \ & \ \ \ \ \ ...\ \ \ \ \ &  15.7226 &   0.2460 &   1.6739 \\
 01\ 31\ 54.593 & -13\ 37\ 04.27 &  19.4932 &   0.2552 &   3.4488 & \ \ \ \ \ ...\ \ \ \ \ & \ \ \ \ \ ...\ \ \ \ \ & \ \ \ \ \ ...\ \ \ \ \ \\
 01\ 31\ 55.124 & -13\ 37\ 04.70 &  20.6571 &   1.0807 &   4.3701 &  16.4715 &   0.6599 &   4.1874 \\
 01\ 31\ 55.177 & -13\ 36\ 57.94 &  22.3937 &   3.2107 &   7.4987 &  17.3259 &   1.1985 &   7.6776 \\
 01\ 31\ 51.069 & -13\ 36\ 27.59 &  20.2655 &   0.2876 &   1.7283 &  17.4428 &   0.4174 &   1.2552 \\
 01\ 31\ 52.292 & -13\ 36\ 58.27 &  20.5305 &   1.8861 &   4.4878 &  16.8956 &   1.5686 &   5.3478 \\
 01\ 31\ 51.793 & -13\ 36\ 54.83 &  19.6565 &   0.3473 &   4.8037 & \ \ \ \ \ ...\ \ \ \ \ & \ \ \ \ \ ...\ \ \ \ \ & \ \ \ \ \ ...\ \ \ \ \ \\
 01\ 31\ 52.522 & -13\ 36\ 27.50 &  20.5520 &   0.6928 &   2.0695 &  16.9357 &   0.5250 &   2.6691 \\
 01\ 31\ 53.251 & -13\ 36\ 44.19 &  19.1673 &   0.2396 &   7.7844 & \ \ \ \ \ ...\ \ \ \ \ & \ \ \ \ \ ...\ \ \ \ \ & \ \ \ \ \ ...\ \ \ \ \ \\
 01\ 31\ 52.868 & -13\ 36\ 35.24 & \ \ \ \ \ ...\ \ \ \ \ & \ \ \ \ \ ...\ \ \ \ \ & \ \ \ \ \ ...\ \ \ \ \ &  18.2153 &   0.8581 &   9.9032 \\
 01\ 31\ 52.570 & -13\ 36\ 44.18 & \ \ \ \ \ ...\ \ \ \ \ & \ \ \ \ \ ...\ \ \ \ \ & \ \ \ \ \ ...\ \ \ \ \ &  16.7582 &   0.6264 &   5.8563 \\
 01\ 31\ 52.526 & -13\ 36\ 40.46 &  22.0511 &   6.7923 &   2.9558 &  20.3538 &  21.9563 &   5.2298 \\
 01\ 31\ 50.965 & -13\ 36\ 49.60 &  22.0171 &   1.9828 &   7.1752 &  16.8061 &   0.5905 &   3.6298 \\
 01\ 31\ 56.222 & -13\ 36\ 46.75 &  21.9203 &   2.6347 &   3.7004 &  17.7281 &   1.2374 &   2.1518 \\
 01\ 31\ 48.638 & -13\ 36\ 46.60 &  19.7038 &   0.2758 &   4.3260 &  17.2784 &   0.3859 &   3.8515 \\
 01\ 31\ 51.318 & -13\ 36\ 56.77 &  21.9856 &   4.3616 &   9.6712 &  16.9624 &   1.5630 &   4.9491 \\
 01\ 31\ 57.703 & -13\ 36\ 43.59 &  19.8645 &   0.2750 &   2.4002 &  16.8334 &   0.2815 &   1.5780 \\
 01\ 31\ 57.325 & -13\ 36\ 32.77 &  20.6240 &   0.5081 &   1.4938 &  17.9624 &   0.5168 &   0.8189 \\
 01\ 31\ 48.709 & -13\ 36\ 25.01 &  19.7810 &   0.4451 &   1.4137 &  16.1749 &   0.3776 &   2.2878 \\
 01\ 31\ 51.259 & -13\ 36\ 20.88 &  20.1945 &   0.3756 &   1.8985 &  17.4433 &   0.4172 &   1.2512 \\
 01\ 31\ 53.329 & -13\ 36\ 31.32 &  19.2165 &   0.5749 &   6.5730 &  16.0117 &   0.5300 &   4.4363 \\
 01\ 31\ 53.837 & -13\ 36\ 13.04 &  19.4277 &   0.7632 &   3.8471 &  15.6959 &   0.6156 &   4.1877 \\
 01\ 31\ 49.361 & -13\ 36\ 06.45 &  19.4235 &   0.3293 &   1.9263 &  16.4016 &   0.3427 &   1.5411 \\
 01\ 31\ 49.825 & -13\ 36\ 11.24 &  19.6579 &   0.4963 &   6.3710 &  16.3664 &   0.4081 &   2.4116 \\
 01\ 31\ 52.968 & -13\ 36\ 22.32 &  19.0905 &   0.3023 &   1.9196 &  15.7089 &   0.3013 &   5.0108 \\
 01\ 31\ 50.863 & -13\ 36\ 03.80 &  20.0735 &   1.0809 &   3.7781 &  16.5416 &   0.9476 &   3.5447 \\
 01\ 31\ 50.312 & -13\ 36\ 01.40 & \ \ \ \ \ ...\ \ \ \ \ & \ \ \ \ \ ...\ \ \ \ \ & \ \ \ \ \ ...\ \ \ \ \ & \ \ \ \ \ ...\ \ \ \ \ & \ \ \ \ \ ...\ \ \ \ \ & \ \ \ \ \ ...\ \ \ \ \ \\
 01\ 31\ 53.668 & -13\ 36\ 03.78 &  21.0414 &   0.4024 &   1.5757 & \ \ \ \ \ ...\ \ \ \ \ & \ \ \ \ \ ...\ \ \ \ \ & \ \ \ \ \ ...\ \ \ \ \ \\
 01\ 31\ 50.356 & -13\ 35\ 52.89 &  19.0362 &   0.2349 &   0.3000 &  16.4482 &   0.2815 &   0.3000 \\
 01\ 31\ 56.881 & -13\ 35\ 51.56 &  20.5237 &   0.4220 &   2.0263 &  17.8736 &   0.4903 &   1.0736 \\
 01\ 31\ 22.104 & -13\ 24\ 37.48 &  19.6973 &   0.5614 &   2.3988 & \ \ \ \ \ ...\ \ \ \ \ & \ \ \ \ \ ...\ \ \ \ \ & \ \ \ \ \ ...\ \ \ \ \ \\
 01\ 31\ 25.470 & -13\ 25\ 15.52 &  21.4579 &   0.5904 &   4.6591 &  17.0320 &   0.3165 &   6.0199 \\
 01\ 31\ 28.920 & -13\ 25\ 19.18 &  19.6969 &   0.2542 &   0.7850 &  16.8525 &   0.3937 &   5.8122 \\
 01\ 31\ 33.649 & -13\ 25\ 32.56 &  20.5194 &   1.1055 &   2.9986 &  17.3696 &   1.1272 &   3.3754 \\
 01\ 31\ 25.207 & -13\ 26\ 33.00 & \ \ \ \ \ ...\ \ \ \ \ & \ \ \ \ \ ...\ \ \ \ \ & \ \ \ \ \ ...\ \ \ \ \ & \ \ \ \ \ ...\ \ \ \ \ & \ \ \ \ \ ...\ \ \ \ \ & \ \ \ \ \ ...\ \ \ \ \ \\
 01\ 31\ 24.652 & -13\ 26\ 08.03 &  21.0140 &   2.0250 &   6.1824 &  17.2039 &   1.5147 &   8.1053 \\
 01\ 31\ 33.012 & -13\ 26\ 46.86 &  20.7910 &   0.7239 &   3.0107 &  17.6311 &   0.6161 &   1.7465 \\
 01\ 32\ 00.956 & -13\ 27\ 05.96 &  18.5969 &   0.3669 &   3.8551 &  16.0508 &   0.4125 &   1.2808 \\
 01\ 31\ 45.290 & -13\ 27\ 14.58 &  19.2001 &   0.6152 &   4.1630 &  15.3983 &   0.5332 &   6.6005 \\
 01\ 31\ 22.896 & -13\ 27\ 32.46 &  19.7890 &   0.2410 &   2.9062 &  16.7471 &   0.3168 &   2.9581 \\
 01\ 31\ 40.989 & -13\ 27\ 22.20 &  21.4567 &   1.4440 &   2.9251 &  17.9830 &   1.0992 &   3.1077 \\
 01\ 31\ 32.956 & -13\ 27\ 25.09 &  20.3794 &   0.6837 &   2.7252 &  17.2724 &   0.6821 &   4.1649 \\
 01\ 31\ 30.360 & -13\ 27\ 08.93 &  21.5145 &   2.1049 &   3.1130 &  19.5043 &   4.4167 &   6.1853 \\
 01\ 31\ 57.443 & -13\ 28\ 15.91 &  20.2850 &   0.3420 &   3.9310 &  16.9001 &   0.2943 &   1.8631 \\
 01\ 31\ 35.616 & -13\ 28\ 16.12 &  22.6539 &   2.3584 &   6.4603 &  20.3706 &   3.8516 &   8.9163 \\
 01\ 31\ 21.567 & -13\ 28\ 27.36 & \ \ \ \ \ ...\ \ \ \ \ & \ \ \ \ \ ...\ \ \ \ \ & \ \ \ \ \ ...\ \ \ \ \ & \ \ \ \ \ ...\ \ \ \ \ & \ \ \ \ \ ...\ \ \ \ \ & \ \ \ \ \ ...\ \ \ \ \ \\
 01\ 31\ 19.450 & -13\ 28\ 16.76 &  18.3187 &   0.2393 &   5.4436 & \ \ \ \ \ ...\ \ \ \ \ & \ \ \ \ \ ...\ \ \ \ \ & \ \ \ \ \ ...\ \ \ \ \ \\
 01\ 31\ 46.413 & -13\ 28\ 29.23 &  21.6343 &   1.2590 &   2.7800 &  19.2229 &   1.9042 &   4.0375 \\
 01\ 31\ 27.472 & -13\ 28\ 36.00 &  18.6471 &   0.3872 &   3.1090 & \ \ \ \ \ ...\ \ \ \ \ & \ \ \ \ \ ...\ \ \ \ \ & \ \ \ \ \ ...\ \ \ \ \ \\
 01\ 31\ 19.532 & -13\ 28\ 46.75 &  20.5445 &   0.3773 &   3.9486 &  16.6523 &   0.2603 &   4.8619 \\
 01\ 31\ 34.200 & -13\ 28\ 51.22 &  19.7480 &   0.5969 &   3.1140 &  16.6719 &   0.5749 &   3.4278 \\
 01\ 31\ 21.362 & -13\ 28\ 48.04 &  20.2568 &   0.4839 &   2.3031 &  16.5964 &   0.3630 &   2.8755 \\
 01\ 31\ 28.102 & -13\ 28\ 54.14 &  18.5654 &   0.2678 &   4.6348 &  16.3346 &   0.4461 &   7.1639 \\
 01\ 31\ 28.097 & -13\ 28\ 57.45 &  21.1347 &   1.3199 &   6.0171 &  17.0114 &   0.8716 &   8.5200 \\

            \noalign{\smallskip}	     		    
            \hline			    		    
         \end{array}
     $$ 
}
         \end{table}

\subsection{Kromendy relation and Photometric Plane}
\label{sec:62}

ETGs define a remarkable correlation between the
effective radius r$\mathrm{_e}$ and the mean surface brightness
$\mathrm{<\mu_e>}$, known as Kormendy relation (KR, Kormendy
\citeyear{kor77}): $$\mathrm{<\mu_e>} = \ \ \alpha \ \ + \ \ \beta \ \
\mathrm{log r_e} \ \ \ .$$

In order to study the Kormendy relation (hereafter KR), we selected the
population of spheroids on the basis of the shape of the light
profile, as parameterised by the Sersic index $n$. We classified as
spheroids the galaxies with $n>2$, corresponding to objects within a
bulge fraction greater than $\sim$ 20\% (see Saglia et
al. \citeyear{sag97}; van Dokkum et al. \citeyear{van98}). Moreover,
17 and 54 objects in R and K-band respectively, with small radii were
excluded by selecting only galaxies with $r_e$ greater than 1 pixel.

Again, we point out that the present samples constitute the largest
data set of galaxies belonging to one cluster for which the KR is
obtained. In fact 240 and 227 galaxies are identified as spheroids in
R and K band respectively, with spectroscopic redshift or photometric
redshift in the range $0.14-0.26$ (see above).

Since selection effects can strongly affect the estimate of the KR
coefficients (see Ziegler et al. \citeyear{zie99}), we fitted the
$\mathrm{log r_e}$-$\mathrm{<\mu_e>}$ sequences by introducing a
modified least-squares (MLS) procedure (see La Barbera et
al. \citeyear{lab03b}), which corrects the bias due to the different
completeness cuts in magnitude of each sample. The fitting
coefficients were derived by applying the bisector regression (see
Akritas \& Bershady \citeyear{akr96}).

\begin{figure}
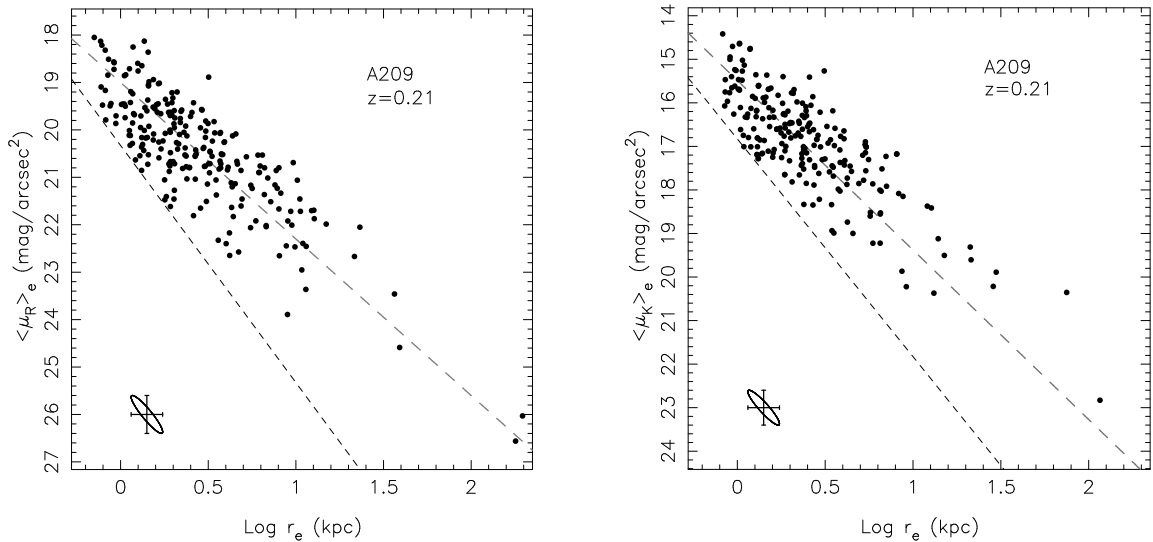

\hbox{
{\resizebox{7cm}{!}{\includegraphics{fig14a.ps}}}
\hfill
\hspace{1cm}
{\resizebox{7cm}{!}{\includegraphics{fig14b.ps}}}
}
\caption{Kormendy relations for ABCG\,209 in R (right panel) and K
band. MLS$_{\mathrm{logr_e}}$ is represented by long dashed
lines. The short dashed line indicate the cuts in total
magnitude. The correlation of the uncertainties on
${\mathrm{logr_e}}$ and ${\mathrm{<\mu_e>}}$ is shown by the
ellipses (1 $\sigma$ confidence contours) in the lower left of each
panel.}
\label{KR}
\end{figure}

Figure~\ref{KR} shows the KR for R (left panel) and K band (right
panel), reporting also the correlation between $\delta \mathrm{log
r_e}$ and $\delta \mathrm{<\mu_e>}$, with the typical values of the
uncertainties on the effective parameters. We obtain zero points
$\alpha= 19.19\pm0.07$ and $\alpha= 15.68\pm0.07$, and slopes $\beta=
3.07\pm0.16$ and $\beta=3.26\pm0.19$ for R and K bands, respectively.

The difference in the zero points is fully consistent with the
colour R-K of galaxies, according to the CM relation
(Fig.~\ref{figCM}).

The slope of the KR in R band is fully consistent with that derived in
La Barbera et al. (\citeyear{lab03b}) for a smaller (N=81) and
shallower (R$<$20.1) sample of galaxies in ABGC\,209, thus confirming
that the R-band KR is invariant in the redshift range from z=0.023 to
z=0.64. For what the K-band KR is concerned, our slope is fully
consistent with that found by La Barbera et al. (\citeyear{lab04}) for
the galaxies in the cluster A2163B at z$\sim$0.2.

The structural properties of early-type galaxies define a plane in the
three-dimensional space $(\mathrm{log r_e},\mathrm{log
n},\mathrm{<\mu_e>})$, the so-called PHP (see Graham
\citeyear{gra02} and references therein):

$$\mathrm{log r_e} = \ \ a \ \ \mathrm{log n} + b \ \ \mathrm{<\mu_e>}
+ c \ \ .$$

We derived coefficients of the PHP by using the corrected orthogonal
weighted least-squares fit (La Barbera et al. \citeyear{lab05}). This
method, treating equally all the variables, gives much more robust
estimates for the coefficients of the PHP and is less sensitive than
others to selection effects.

\begin{figure}
\includegraphics[angle=-90,width=1.0\textwidth]{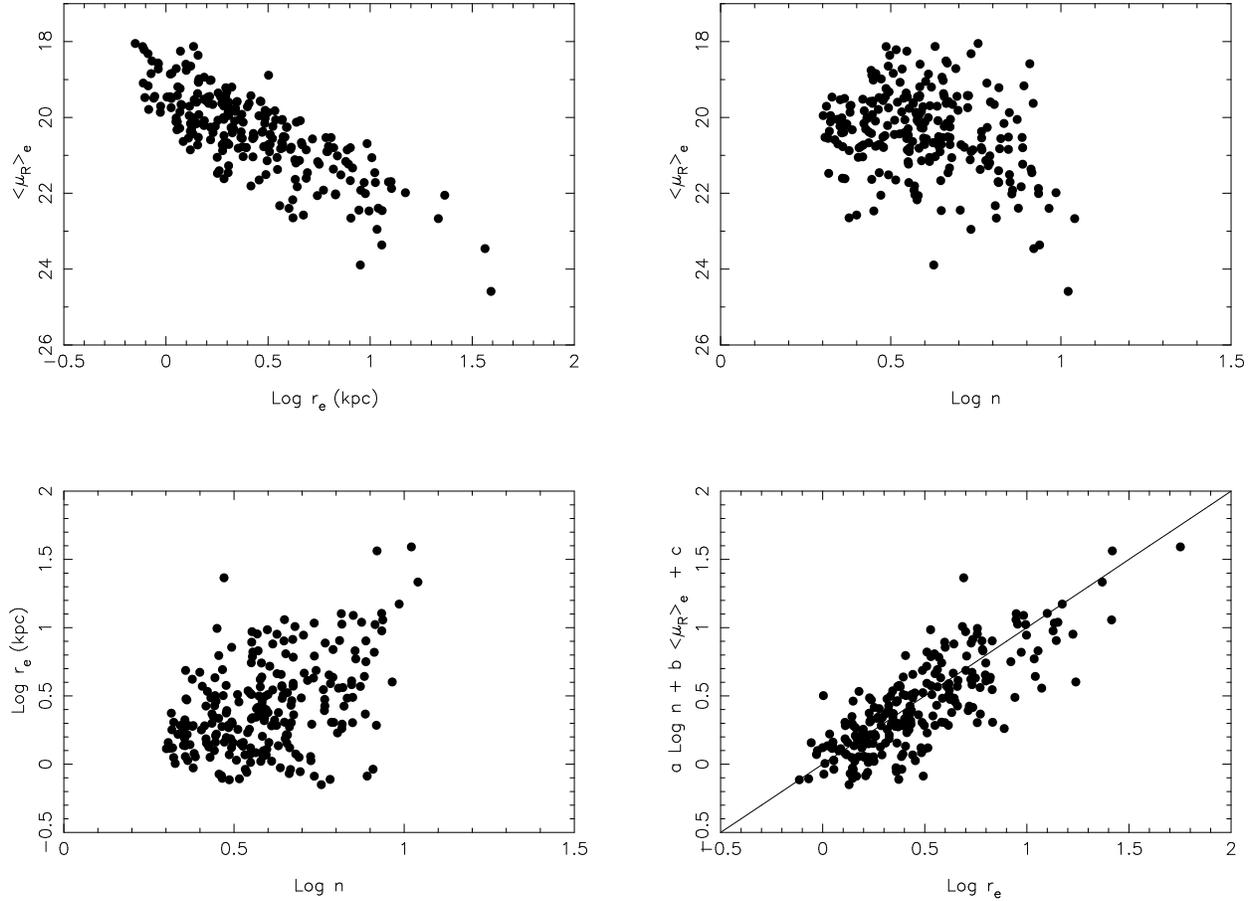}

\caption{Photometric Plane of ABCG\,209 in the R band. The upper
panels and the lower--left panel show the $log r_e-<\mu_e>$, $log
n-<\mu_e>$ and $log n-log r_e$ projections of the PHP, respectively.
The lower-right panel shows the edge-on view of the R-band Photometric
Plane. The plots show the N= 240 galaxies of the R-band sample. }

\label{PHP_R}
\end{figure}

\begin{figure}
\includegraphics[angle=-90,width=1.0\textwidth]{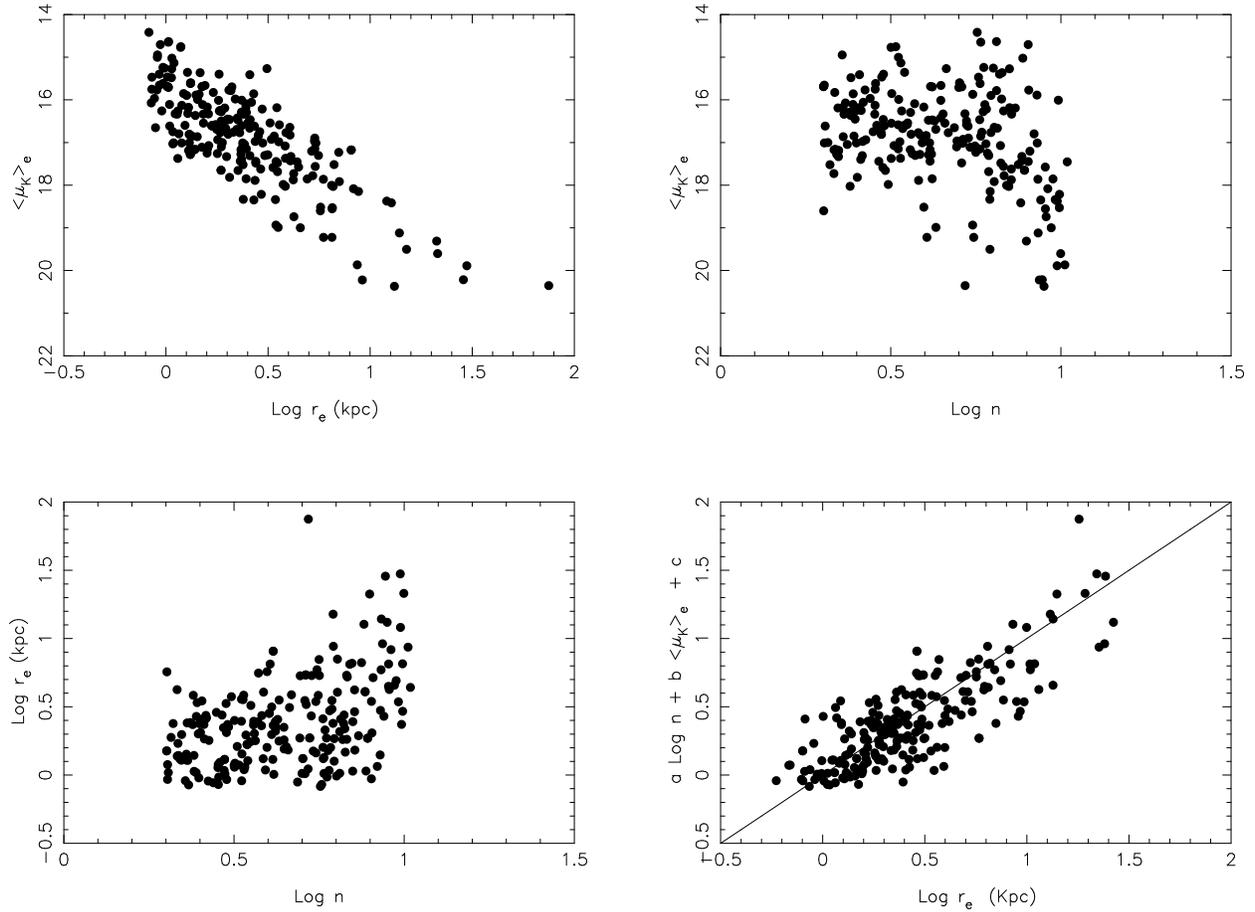}

\caption{Photometric Plane of ABCG\,209 in the K band. The upper
panels and the lower--left panel show the $log r_e-<\mu_e>$, $log
n-<\mu_e>$ and $log n-log r_e$ projections of the PHP, respectively.
The lower-right panel shows the edge-on view of the K-band Photometric
Plane. The plots show the N= 227 galaxies of the K-band sample. }

\label{PHP_K}
\end{figure}

The distribution of galaxies in the space of structural parameters is
shown in Figs.~\ref{PHP_R} and \ref{PHP_K}, for R and K band
respectively. We show the distribution of galaxies in the $log
r_e-<\mu_e>$, $log n-<\mu_e>$ and $log n-log r_e$ planes, and an edge
projection of the plane. Galaxies follow a well-defined PHP at z= 0.2,
with Sersic indices that increase towards lower surface brightness
values and larger effective radii.

In Table~\ref{tab_PHP} we report the coefficient of the PHP for R and
K band. The coefficients of the R- and K-band PHP are consistent (at
1$\sigma$) with those obtained by La Barbera et al. (\citeyear{lab05})
for the cluster MS 1008 at z$\sim$0.3 and by Graham et
al. (\citeyear{lab05}) for nearby clusters, confirming that PHP seems
to be independent of redshift. 

\begin{table}
        \caption[]{Coefficients of the PHP.}
         \label{tab_PHP}
           $$ 
           \begin{array}{c c c c c}
            \hline
            \noalign{\smallskip}
            \hline
            \noalign{\smallskip}

\mathrm{Band} & \mathrm{a}  & \mathrm{b}  & \mathrm{c}  &
            \mathrm{\sigma_{Log r_e}} \\
            \hline
            \noalign{\smallskip}   

R  & 0.964\pm0.130 &  0.209\pm0.020 & -4.376\pm0.401 &  0.162\pm0.014\\
K  & 0.711\pm0.098 &  0.227\pm0.022 & -3.871\pm0.404 &  0.167\pm0.016\\

              \noalign{\smallskip}			    
            \hline					    
            \noalign{\smallskip}			    
            \hline					    
         \end{array}
   $$ 
         \end{table}

\subsection{The intrinsic dispersion of the PHP}
\label{sec:63}

In Table~\ref{tab_PHP} we showed that the PHP has a small intrinsic
dispersion ($\sigma_{Log r_e} =0.16-0.17$) both in R and K band.  La
Barbera et al. \citeyear{lab05} pointed out that stellar populations
can be the origin of the dispersion about the plane. However they
have shown that the residuals of the I-K versus K colour-magnitude
relation do not correlate with the residuals about the PHP.

The dispersion about the PHP could be due to the combined effect of
age, metallicity, and also $\alpha$/Fe enhancement, that cannot be
distinguished only with the colour-magnitude relation. For this reason it
is crucial to use line strength to address this issue.

In order to distinguish between the contribution of age, metallicity
and $\alpha$/Fe enhancement, we compared the line indices obtained in
Mercurio et al. (\citeyear{mer04}) with the residuals of PHP.  We
analyse a subsample of 83 and 91 ETGs, in R and K band respectively,
for which we have these measurements. In particular we use
H$_{\beta}$, [MgFe]$^\prime$ (Thomas et al. \citeyear{tho03}) and
Mgb/$<$Fe$>$, as the best indicators of age, metallicity and
$\alpha$/Fe enhancement, respectively (see also Puzia at
al. \citeyear{puz05}).

\begin{figure}
\begin{center}
\includegraphics[angle=-90,width=0.6\textwidth]{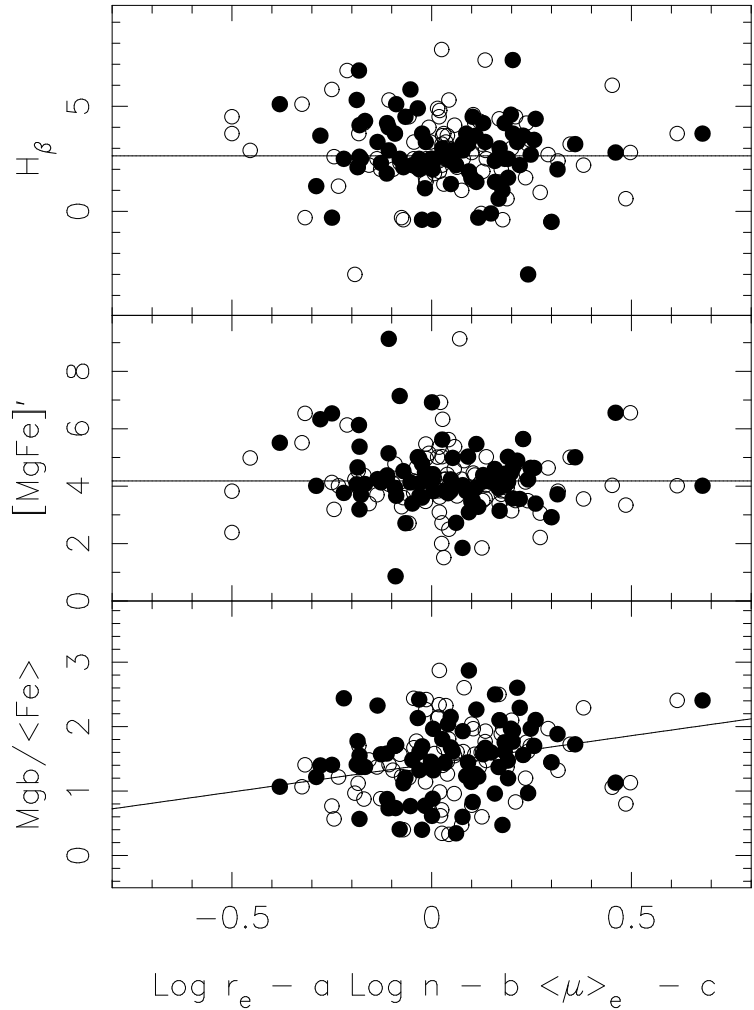}
\end{center}
\caption{Correlation between PHP residuals and equivalent width of the
H$_\beta$ (upper panel), strength of [MgFe]$^\prime$ (central panel)
and Mgb/$<$Fe$>$ (lower panel) for R (filled circles) and K (open
circles) band. Continuous lines represent the least-square fit on the
data.}
\label{PHP_EW}
\end{figure}

Figure~\ref{PHP_EW} shows that the H$_\beta$ (upper panel) and
[MgFe]$^\prime$ indices (central panel) do not correlate with either
the R-band PHP (filled circles) or the K-band PHP (open circles).
The dispersion of the PHP seems instead related to Mgb/$<$Fe$>$
measurements (lower panel). 

We perform a least-square fit on the data, obtaining -0.051$\pm$0.461
and -0.261$\pm$0.705 for the age and metallicity correlation slopes,
while for the $\alpha$/Fe we obtain 0.702$\pm$0.258.

This suggest that the chemical evolution of galaxies could be
responsible for the intrinsic dispersion of the PHP, as argued by La
Barbera et al. ({\citeyear{lab05}).

\section{Summary and Conclusions}
\label{sec:7}

In order to study the internal dynamics of the rich galaxy cluster
ABCG\,209, we have analysed spectra for a total sample of 148 galaxies in
the cluster region, of which 134 are candidate cluster members having $z{\sim}0.209$. In a previous paper we investigated the dynamical
status of this cluster, selecting 112 galaxies belonging to the
cluster (Mercurio et al. \citeyear{mer03a}). We complement the
previously described dataset with medium resolution spectra for 29 new
cluster galaxies. We also re-observe 22 cluster members contained in
the already available data sets. The enlarged data sets allow us to
better distinguish among galaxies belonging to different
substructures, to derive their individual velocity distributions and
to study possible spatial segregation.

We extend the existing optical ($B$, $V$ and $R$ bands) photometry
with $K$-band observations which allow us to estimate reliable
photometric redshifts ($\delta z / 1+z \le 0.07$) for 399 galaxies brighter
than $R$=21 mag in the cluster field and to derive the NIR LF. 

The Schechter function provides a good description of cluster galaxy
counts, with $\alpha=-0.98 \pm 0.15$ and $\rm K^* = 14.80 \pm 0.30$
mag, values consistent with previous studies of NIR luminosity
functions of cluster galaxies at $z{\sim}0.2$ (\citealt{dPE98},
\citealt{AnP00}, \citealt{lab04}).  The only magnitude bin showing a
larger deviation is that at $K=17.5$ mag, where a marginal ($\sim 1.5
\sigma$ significant) deficiency of cluster galaxies is found, that
corresponds to the magnitude bin ($20<R<21$ mag) where Mercurio et
al. (\citeyear{mer03b}) find some indication of a dip in the optical
luminosity functions of ABCG\,209.

ABCG\,209 is characterised by a very high value of the LOSVD:
$\sigma_v=1268^{+93}_{-84}$ \kss, that results in a virial mass of
$M_{vir}=2.95^{+0.80}_{-0.78}\times 10^{15} h^{-1}_{70}$ \msun within
$R_{vir}=3.42 h^{-1}_{70}$ Mpc. The analysis of the velocity
dispersion profile shows that such a high value of $\sigma_v$ is
already reached in the central cluster region suggesting the
possibility that a mixture of clumps at different mean velocities
causes the high value of the velocity dispersion and the virial
mass. In fact, we found that a mixture of three Gaussians is the best
description of the velocity distribution (at $98.4\%$ c.l.). Assigning
the member galaxies to individual groups we found the main clump
(n$_{gal}$=102) at mean redshift z = 0.2090 with a velocity dispersion
of ${\rm\sigma_{v}= 847^{+52}_{-49}}$ \kss and other two groups of 16
galaxies each at z = 0.1988 and z = 0.2172 with ${\rm\sigma_{v} =
323^{+47}_{-58}}$ \kss and ${\rm\sigma_{v}=289^{+57}_{-54} }$ \kss
respectively.

This observational scenario confirms that ABCG\,209 is presently
undergoing strong dynamical evolution with the merging of two or more
subclumps, in agreement with the recent detection of a radio halo
(Giovannini et al. \citeyear{gio06}).  On the other hand, there are
discrepancies between the dynamical, X-ray and lensing analyses. While
Smith et al. (\citeyear{smi05}) and Zhang et al. (\citeyear{zha07})
found no evidence for multimodality in the dark matter distribution
and subclumps in the XMM data respectively, an irregular X-ray
morphology was identified by Rizza et al. (\citeyear{riz98}) in ROSAT
data and by Mercurio et al. \citeyear{mer03a} (see their Fig. 13) and
Smith et al. \citeyear{smi05} (see their Fig. 6) in Chandra
observations.  In particular, the Mercurio et al. (\citeyear{mer03a})
analysis recognised two significant substructures. The principal one
centred on the cD galaxy and the second about $50{\arcsec}$ East of
the cD, this secondary clump being coincident with the Eastern clump
detected by Rizza et al. (\citeyear{riz98}).  Since these two detected
clumps are close to each other, probably the absence of evident
subclumps in the XMM-Newton data is due to the limited spatial
resolution of this instrument.

Smith et al. (\citeyear{smi05}) suggest that this discrepancy could be
related to time evolution and that sufficient time has elapsed after
the merger. However, as also pointed out by Smith et
al. (\citeyear{smi05}), numerical simulations suggest that both gas
dynamics and substructure in the dark matter distribution may persist
as long as ${\sim}5$ Gyr after a cluster merger, and that the
relaxation time for these two matter components may be
comparable. Moreover, the detection of a radio halo (Giovannini et
al. \citeyear{gio06}) supports a recent merging event. In fact
numerical simulation indicate that the mergers could generate strong
fluid turbulence, supplying energy to the electrons, which then
radiate in the interstellar medium of radio halos, but the time during
which the process is effective is ${\sim}10^8$ yr (Feretti
\citeyear{fer07}).

In independent lensing analyses of the optical CFHT images,
Paulin-Henriksson et al. (\citeyear{pau07}) and Bardeau et
al. (\citeyear{bar07}) measured virial masses of $M_{200} =
7.7^{+4.3}_{-2.7} \times 10^{14} $ \msun and $M_{200} = 7.2\pm 2.0
\times 10^{14}$ \msun and scaled velocity dispersions
${\rm\sigma_v=924\pm84}$ km s$^{-1}$ and ${\rm\sigma_v=813\pm70}$ km
s$^{-1}$ respectively, which are consistent with the value obtained
for the main clump from the dynamical analysis.  For this reason we
suggest that the observed weak lensing masses are associated with the
dark matter halo of the most prominent clump of ABCG\,209. The clumps
are not spatially segregated so they would not be separable in the
dark matter maps which measure only the projected mass distribution,
and the combined contribution of the secondary clumps is expected to
be much lower than that from the primary one.

We have derived structural parameters of galaxies in the R and K bands
in order to obtain the Kormendy relation and the Photometric Plane for
ABCG\,209, based on a total sample of 240 and 227 spheroids in optical
and NIR, respectively.  Spheroids define a tight sequence in the plane
of the effective parameters with a slope $\beta \sim 3$ in both
bands. This slope of the KR is fully consistent with that derived in
La Barbera et al. (\citeyear{lab03b}) for a smaller (N=81) and
shallower (R$<$20.1) sample of galaxies in ABGC\,209, thus confirming
that the R-band KR is invariant in the redshift range from z=0.023 to
z=0.64. For what the K-band KR is concerned, our slope is fully
consistent with that found by La Barbera et al. (\citeyear{lab04}) for
the galaxies in the cluster A2163B at z$\sim$0.2.

The cluster ETGs at $z{\sim}0.2$ follow a tight correlation among
$\mathrm{\log r_e}$, $<{\mu}>_e$ and log n, with an intrinsic
dispersion of $\sim$0.17 dex in both optical and NIR wavebands. This
dispersion is fully consistent with that found by La Barbera et
al. (\citeyear{lab05}) for the cluster MS1008 at $z{\sim} 0.3$ and
Graham (\citeyear{gra02}) for ETGs in nearby clusters.

In order to investigate the origin of the intrinsic scatter of the PHP
we have analysed the PHP residuals versus line-strength indices.  We
compare the line indices obtained in Mercurio et
al. (\citeyear{mer04}) with the PHP residuals of member galaxies for a
subsample of 83 and 91 ETGs, in R and K band respectively.  In
particular we use H$_{\beta}$, [MgFe]$^\prime$ (Thomas et
al. \citeyear{tho03}) and Mgb/$<$Fe$>$, as the best indicators of age,
metallicity and $\alpha$/Fe enhancement, respectively (Thomas et
al. \citeyear{tho03}, Puzia at al. \citeyear{puz05}).  The PHP
residuals do not correlate with age and metallicity, while there is a
correlation with $\alpha$/Fe enhancement (Fig.~\ref{PHP_EW}). This
could imply that the scatter of the PHP is due to variations in
stellar population parameters, in particular it could be due to
variations in the chemical evolution of early-type galaxies brought
about by their merging histories.

\section*{Acknowledgments}   

AM is funded by the INAF-OAC.

\label{lastpage}
\end{document}